\documentclass[%
 reprint,
superscriptaddress,
 amsmath,amssymb,
 aps,
]{revtex4-2}
\usepackage[normalem]{ulem}
\usepackage{graphicx}% Include figure files
\usepackage{dcolumn}% Align table columns on decimal point
\usepackage{cancel}
\usepackage{bm}% bold math
\usepackage{orcidlink}
\usepackage{comment}
\usepackage{microtype}
\usepackage{xcolor}
\colorlet{RED}{red}
\usepackage{soul}
\usepackage{accents}
\usepackage{verbatim}
\newcommand{\mbX}{\mathbf{X}}

\newcommand{\mbW}{\mathbf{W}}
\newcommand{\mbx}{\mathbf{x}}
\newcommand{\mby}{\mathbf{y}}

\newcommand{\mbf}{\mathbf{f}}
\newcommand{\mbF}{\mathbf{F}}
\newcommand{\mbg}{\mathbf{g}}

\newcommand{\R}{\mathbb{R}}
\newcommand{\LL}{\mathcal{L}}
\newcommand{\LLd}{\mathcal{L}^\dagger}
\newcommand*\diff{\mathop{}\!\mathrm{d}}
\newcommand{\tbar}{\overline{T}} 
\def\given{\:|\:}

\newcommand{\e}[1]{eq.~(\ref{eq:#1})}
\newcommand{\be}{\begin{equation}}
\newcommand{\ee}{\end{equation}}
\newcommand{\lr}[1]{\left\langle #1 \right\rangle}

\newcommand{\Tbar}{\overline{T}}

\newcommand\numberthis{\addtocounter{equation}{1}\tag{\theequation}}

\begin{document}

\preprint{APS/123-QED}

\title{Generalized dynamical phase reduction for stochastic oscillators}% Force line breaks with \\

\author{Pierre Houzelstein \orcidlink{0009-0000-0054-8066}}
\homepage{pierre.houzelstein@ens.psl.eu}
\affiliation{Group for Neural Theory, LNC2 INSERM U960, DEC, École Normale Supérieure \\ PSL University, Paris, France}
\author{Peter J.~Thomas}
\affiliation{Department of Mathematics, Applied Mathematics and Statistics, Case Western Reserve University,\\ Cleveland, Ohio, United States of America}
\author{Benjamin Lindner}
\affiliation{ Bernstein Center for Computational Neuroscience Berlin, Berlin 10115, Germany\\
 Department of Physics, Humboldt Universität zu Berlin, Berlin D-12489, Germany}
 \author{Boris Gutkin}
\affiliation{Group for Neural Theory, LNC2 INSERM U960, DEC, École Normale Supérieure \\ PSL University, Paris, France}
\author{Alberto Pérez-Cervera}
\homepage{albert.prz.crv@gmail.com}
\affiliation{Universitat Politècnica de Catalunya, Barcelona, Spain
}

\begin{comment}
\date{\today}% It is always \today, today,
             %  but any date may be explicitly specified
\end{comment}

\begin{abstract}
Phase reduction is an important tool for studying coupled and driven oscillators. 
The question of how to generalize phase reduction to stochastic oscillators remains actively debated. In this work, we propose a method to derive a self-contained stochastic phase equation of the form
$\diff \phi = a(\phi)\diff t + \sqrt{2D(\phi)}\,\diff W(t)$ that is valid not only for noise-perturbed limit cycles, but also for noise-induced oscillations. We show that our reduction captures the asymptotic statistics of qualitatively different stochastic oscillators, and use it to infer their phase-response properties. 
\end{abstract}

%\keywords{Suggested keywords}%Use showkeys class option if keyword
%display desired
\maketitle

%\tableofcontents

\section{Introduction}

Oscillatory behaviour is an ubiquitous phenomenon in physical, biological, chemical and engineering systems \cite{PikRos03}. A powerful way of approaching oscillations is by means of a phase variable. In a purely deterministic system
\begin{equation}\label{eq:sys-x}
\dot{\mbx} = \mathbf{F}(\mbx), \qquad \mbx \in \mathbb{R}^n
\end{equation}
oscillatory behaviour corresponds to stable $T$-periodic solutions of system (\ref{eq:sys-x}) around the attractor of the dynamics: the limit cycle (LC), which we denote as $\Gamma$. Typically, the existence of the attractor is used to provide a simpler description of the oscillatory dynamics. Namely, one parameterizes the LC, which is a closed curve in the phase space, by means of an angular \textit{phase} variable $\vartheta$ such that $\Gamma = \{ \mbx ~|~ \mbx = \gamma(\vartheta)\}$. Assuming the solutions are asymptotically close to the limit cycle, the parametrization $\gamma(\vartheta)$ allows to study the system $\eqref{eq:sys-x}$ by means of the \textit{phase reduction}
\begin{equation}\label{eq:det-phase}
    \diff\vartheta = \frac{2\pi}{T}\diff t,
\end{equation}
which is a one-dimensional description of the periodic dynamics. This phase reduction approach is a well-known method to study complex oscillatory phenomena, such as response to perturbations, phase locking or synchronization \cite{HopIzh97, Win01}.

Since real-world systems are often intrinsically fluctuating and noisy, it is natural to aim to extend the phase reduction framework to stochastic oscillators. In principle, a meaningful \textit{stochastic phase reduction} should provide a level of understanding of the dynamics similar to the deterministic case, while incorporating the noisy component observed in  %the real
realistic oscillations. 

A  first approach to this question is to consider the noise as a weak perturbation of the LC oscillator \cite{ErmBev11}.
In this case, using a perturbative approach, one can describe the stochastic system by means of the deterministic phase \cite{YosAra08, TerNak09, ErmBev11}. 
Alternatively, extensions of phase reduction to stochastic systems based on variational methods have been proposed \cite{BreMac18, Mac23}.
However, perturbative and variational LC approaches both require the existence of an underlying LC. 
Thus, they have trouble generalizing over the important cases when the addition of noise to a non oscillatory deterministic system leads to noise-induced oscillations \cite{LinGar04}. 

Therefore, a fundamental challenge for building a general stochastic phase reduction is to define a phase observable that does not require the existence of an underlying LC, and that is applicable in the wide range of contexts in which LC and noise-induced oscillations can emerge. 
Overcoming this challenge in a successful way requires going back to the phase definition itself and updating it. 
The deterministic phase is defined in terms of two equivalent notions: either in terms of Poincaré
sections, or of the system’s asymptotic behaviour \cite{Guc75}. 
During the last decade, these two notions of phase have been extended to stochastic oscillators. 
Ten years ago, Schwabedal and Pikovsky \cite{SchPik13} found the natural way of extending Poincaré's approach to noisy oscillators. To this end, they constructed a system of isochrons (curves of ``equal timing") with the \emph{mean return time property}, namely, that the \emph{average} time it would take a trajectory to complete one oscillation and return to some point on the original isochron should equal the mean period of the oscillator, a criterion that can be also related to the solution of a partial differential equation \cite{CaoLin20}. As an alternative to the mean-return-time phase, Thomas and Lindner proposed that a meaningful phase observable (which they denoted as the ``stochastic asymptotic phase") can be extracted from the asymptotic behaviour of the conditional density \cite{ThoLin14}. 

However, while these two notions of phase solve the problem of finding a phase observable that applies to the many different mechanisms generating stochastic oscillations, a general method for finding a self-contained one-dimensional Markovian phase equation of the form
\begin{equation}\label{eq:genReduction}
	\diff\phi = a_\phi(\phi)\diff t + \sqrt{2D_\phi(\phi)}\,\diff W_\phi(t),
\end{equation}
that approximates the full process, with both $a_\phi$ and $D_\phi$ smooth and periodic in $\phi$, is still missing. While there have been different attempts in the past, they were built ad-hoc for specific classes of stochastic oscillators \cite{PowLon21,ZhuKat22,Mac23}. 

In this paper, we aim to fill this gap by developing a generalized reduction procedure: given a phase observable $\phi$, we provide a way to obtain a self-contained phase equation as in  \e{genReduction}.  
We show the generality of our procedure by i) applying it to two different phase observables (the previously mentioned Mean-Return-Time phase and the stochastic asymptotic phase) and ii) finding self-contained phase equations of qualitatively different noisy oscillators.

 Our paper is organised as follows. 
 In sec.~\ref{sec:sec-2}, we introduce the mathematical background, which relies on the spectral decomposition of the Kolmogorov backwards operator $\LLd$. In sec.~\ref{sec:phase-maps} we introduce two different phase observables defined via $\LLd$. Next, in sec.~\ref{sec:sec-3}, we introduce the main result of this work: the stochastic phase reduction procedure. In sec.~\ref{sec:sec-4}, we introduce two systems in which oscillations emerge from different mechanisms, and to which we apply our framework. In sec.~\ref{sec:sec-5}, we define the asymptotic statistics  we use to evaluate the quality of our reduction procedure. In sec.~\ref{sec:sec-6}, we show a direct application of our framework: predicting the phase-dependent response of the reduced oscillator to an external perturbation. Next, in sec.~\ref{sec:extension-high-dim} we show how our results extend beyond the planar case. We end with a discussion of the results in sec.~\ref{sec:sec-discussion}.

\section{Theory \& Mathematical Preliminaries}\label{sec:sec-2}

We assume that the stochastic oscillator is described by a multidimensional Markov process with almost surely continuous sample paths.
In particular, we consider
the Itô stochastic differential equation (SDE)
\begin{equation}
		\label{eq:SDE}
		\diff\mbX = \mbf(\mbX)\diff t + \mbg(\mbX)\diff\mbW(t),
\end{equation} 
with $\mbX \in \mathcal{D} \subset \mathbb{R}^n$ the state vector, and where $\mbW \in \mathbb{R}^k$ is a collection of IID Wiener processes with increments $\diff\mbW(t)$.

Instead of studying system \e{SDE} by means of individual realizations (a \textit{pathwise} approach), we adopt an \textit{ensemble} perspective: we consider a collection of trajectories described by the conditional probability density function $P(\mathbf{x}, t|\mathbf{x_0}, t_0)$. We use the standard convention in which $\mbX$ refers to the random variable, while $\mbx$ refers to the independent argument of the corresponding probability density ($\mbX$ is stochastic whereas $\mbx$ is a deterministic object).

This density dynamics obeys the Kolmogorov forward (Fokker-Planck) and Kolmogorov backward equations \cite{Gar04}:
%(also known as the Fokker-Planck equations) \cite{Gar04}:
\begin{flalign}
        \frac{\partial}{\partial t}P(\mathbf{x}, t|\mathbf{x_0}, t_0)& = \mathcal{L}[P] = - \nabla \cdot \vec{J}(\mbx) \notag \\
        & = -\nabla_{\mathbf{x}} \cdot (\mathbf{f}(\mathbf{x})P) + \boldsymbol{\nabla}^2_{\mathbf{x}}(\mathcal{G}(\mathbf{x})P), \label{eq:kolmogorov-forward} \\
        -\frac{\partial}{\partial t_0}P(\mathbf{x}, t|\mathbf{x_0}, t_0) & = \mathcal{L}^{\dagger}[P] \notag \\
        &= \mathbf{f}^\mathsf{T}(\mathbf{x_0}) \cdot \nabla_\mathbf{x_0} P + \mathcal{G}(\mathbf{x_0})\boldsymbol{\nabla}^2_{\mathbf{x_0}}P \label{eq:kolmogorov-backwards},
\end{flalign}
where $\mathcal{G}=\frac12 \mbg\mbg^\intercal$. In eq.\eqref{eq:kolmogorov-forward}, we introduced $\vec{J}(\mbx)$, the probability current
\begin{equation}
        \vec{J}(\mbx) = \vec{\mathbf{f}}(\mathbf{x})P -  \vec{\boldsymbol{\nabla}}_{\mathbf{x}}(\mathcal{G}(\mathbf{x})P),
\end{equation}
which allows us to naturally introduce boundary conditions for the operator $\LL$ in the domain $\mathcal{D}$ as
\begin{equation}\label{eq:boundary-cond}
    \vec{n} \cdot \vec{J} (\mbx) = 0, \enskip\quad \mbx \in \partial\mathcal{D},
\end{equation}
with $\vec{n}$ the normal vector to the boundary. This ensures the total probability density is preserved.
%
%The boundary conditions for $\LLd$ correspond to the adjoint boundary conditions of the forward operator $\LL$ written in \eqref{eq:boundary-cond}. The so-called adjoint reflecting boundary conditions write as
The boundary conditions for $\LLd$ are adjoint to those of $\LL$ and write as
\begin{equation}
(\mathcal{G}(\mbx)\cdot\vec{\boldsymbol{\nabla}}_{\mathbf{x}} F(\mbx))\cdot \vec{n} = 0.
\end{equation}

The backward operator $\mathcal{L}^\dagger$ in \eqref{eq:kolmogorov-backwards} is known as the generator of the Markov process $\mbX(t)$, and is the infinitesimal generator of the family of stochastic Koopman operators associated with \e{SDE} \cite{CrnMac20, KluNus20}. If we define the Koopman semigroup of operators $\mathcal{K}^{\Delta t}$ acting on a real valued observable $F(\mbx)$ of system \eqref{eq:SDE} such that 
\begin{equation}
\mathcal{K}^{\Delta t}[F(\mbx(t))] = \langle F(\mbx(t+\Delta t) \rangle,\end{equation}
then \cite{KluNus20}
\begin{equation}
\mathcal{L}^\dagger[F] = \lim_{\Delta t \to 0}\frac{\mathcal{K}^{\Delta t}[F(\mbx(t))] - F(\mbx(t))}{\Delta t}.
\end{equation}

We assume that the forward ($\mathcal{L}$) and backward ($\mathcal{L}^\dagger$) Kolmogorov operators possess a discrete spectrum with a one-dimensional null space, and eigenvalues $\lambda$ and eigenfunctions $P_\lambda, Q^*_{\lambda}$ satisfying
\begin{equation}
		\label{eq:spectral-eigenfunctions}
		\LL[P_\lambda]=\lambda P_\lambda,\qquad\LLd[Q^*_{\lambda}]=\lambda Q^*_{\lambda}.
\end{equation}
Biorthogonality of the eigenfunctions under the natural inner product follows:
\begin{equation}\label{eq:spectral-decomposition}
        \langle Q_{\lambda'}\given P_{\lambda}\rangle= \int d\mbx\, Q^*_{\lambda'}(\mbx)P_\lambda(\mbx)  =  \delta_{\lambda'\lambda}.
\end{equation}
This relation allows to decompose the conditional probability density as follows \cite{Gar04}: 
for $t>t_0$,
\begin{equation}\label{eq:condDensity}
		P(\mbx,t|\mbx_0,t_0) = P_0(\mbx) + \sum_{\lambda\not=0} e^{\lambda(t-t_0)} P_\lambda(\mbx) Q^*_\lambda(\mbx_0),
\end{equation}
where $P_0$ is the eigenfunction associated with eigenvalue $0$. Properly normalized, it gives the stationary probability density, which in turn defines a stationary density current $\vec{J}_0(\mbx)$. As we are considering stochastic oscillatory systems, which are out of detailed balance, we assume $\vec{J}_0(\mbx)$ to be non-vanishing.

\medskip

According to Itô's chain rule \cite{Gar04, CrnMac20}, for any smooth ($C^2$) observable $F(\mbX)$
\begin{equation}
    dF(\mbX) = \mathcal{L}^\dagger[F(\mbX)] \diff t + \nabla F(\mbX)^\intercal \mbg(\mbX)  \diff \mbW (t).
\end{equation}
Thus, for any stochastic process, the ensemble properties and pathwise realizations of the system are linked through the Kolmogorov backwards operator.

\section{Established phase mappings}\label{sec:phase-maps}

The second-order differential operator $\LLd$ has been used to define two distinct notions of phase reduction for stochastic oscillators, which do not require the existence of an underlying LC: the stochastic asymptotic phase $\Psi(\mbx)$, based on the spectral decomposition of the operator \cite{ThoLin14}, and the Mean--Return-Time phase $\Theta(\mbx)$ \cite{SchPik13}, which is defined in terms of a mean first-passage time problem involving the same operator $\LLd$ \cite{CaoLin20}. 
While these two notions of phase, detailed below, arise from the same operator $\LLd$, they are formally and quantitatively distinct \cite{PerLin22}. Nevertheless, in the case of a system consisting of a LC perturbed by noise, both mappings, $\Psi(\mbx)$ and $\Theta(\mbx)$, converge to the deterministic  asymptotic phase $\vartheta(\mbx)$ in the limit of small noise \cite{CaoLin20, MauMez20, KatZhu21}.

\subsection{The stochastic asymptotic phase}\label{sec:asymp-phase}

Consider system \eqref{eq:SDE} and its associated decomposition of $P(\mbx,t|\mbx_0,t_0)$ \eqref{eq:spectral-decomposition}. 
As proposed in reference \cite{ThoLin14}, if the spectrum of $\LLd$ fulfills the following heuristic conditions:
\begin{enumerate}
\item there exists a nontrivial eigenvalue of $\mathcal{L}^\dagger$ with least negative real part $\lambda_1 = \mu_1 + i\omega_1$, which is complex valued ($\omega_1>0$) and unique;
\item the oscillation is pronounced, i.e.~the \textit{quality factor} $|\omega_1/\mu_1|$ is much larger than 1;
\item all other nontrivial eigenvalues $\lambda'$ are significantly more negative in their real parts, i.e.~$|\Re[\lambda']| \geq  2|\Re[\lambda_1]|$,
\end{enumerate}
then, the approach to the stationary distribution is dominated by a long-lived oscillatory mode $e^{\lambda_1(t-t_0)} P_{\lambda_1}(\mbx) Q^*_{\lambda_1}(\mbx_0) + c.c$, even after all other modes in \e{spectral-decomposition} have decayed \footnote{N.B.~in the Koopman operator literature the term ``mode'' connotes the projections of observables on eigenfunctions. -- I.~Mezić, personal communication.}. 
Appendix~\ref{app:spectra} provides examples of different spectra with such structure.
As discussed in \cite{ThoLin14}, the argument $\Psi(\mathbf{x})$ of the complex backward eigenfunction $Q^*_{\lambda_1}$
\begin{equation}\label{radialQ}
Q^*_{\lambda_1}(\mbx) = u(\mathbf{x})e^{ i\Psi(\mathbf{x})},
\end{equation}
is the natural generalization of the
deterministic asymptotic phase: at large times, and provided the previous set of conditions (denoted in \cite{ThoLin14} as the  \emph{robustly oscillatory} criterion) is met, if one considers the same system at initial time $t\!=\!t_0$ with two different initial conditions $(\mbx(t_0)\!=\!\mbx_1)$ and $(\mbx(t_0)\!=\!\mbx_2)$, the respective probability densities $P(\mby, t|\mbx_1, t_0)$ and  $P(\mby, t|\mbx_2, t_0)$ will decay to the stationary state with an oscillatory offset given by $\Psi(\mbx_1) - \Psi(\mbx_2)$. Thus, $\Psi(\mbx)$ defines level sets, 
\begin{equation}\label{eq:asymp-isos}
    \mathcal{I}_\psi(\mbx) = \{ \mbx ~|~ \Psi(\mbx) = \psi \},
\end{equation}
corresponding to the sets of initial conditions such that the main oscillatory component of their conditional probability densities will evolve in-phase with each other. For this reason, $\Psi(\mbx)$ was denoted in \cite{ThoLin14} as the \textit{ stochastic asymptotic phase}.

Applying the Itô chain rule to this new observable $\Psi(\mbx)$, we extract its evolution law \cite{PerLin22}
\begin{equation}
\begin{aligned}
	\label{eq:psiDyn}
 	  \diff \Psi(\mbX) =& \Big(\omega_1 -  \overbrace{2 \sum_{i,j}\mathcal{G}_{ij}(\mbX) \partial_i \ln(u(\mbX)) \partial_j \Psi(\mbX)}^{\Omega(\mbX)}\Big) \diff t\\ &+ \nabla \Psi(\mbX)^\intercal \mbg(\mbX)  \diff \mbW (t), 
\end{aligned}
\end{equation}
where we introduce the function $\Omega(\mbx)$ to ease notation.

\subsection{The Mean--Return-Time Stochastic Phase}

An alternative definition for the phase of stochastic oscillators was proposed by Schwabedal and Pikovsky in \cite{SchPik13}, who constructed the \textit{Mean-Return-Time phase} in terms of a system of Poincaré sections, which we write $ \ell_\text{MRT}(\varphi)$, with$\enskip 0 \leq \varphi \leq 2\pi $, foliating a domain $\mathcal{R} \subset \mathbb{R}^2$ and possessing a Mean--Return-Time (MRT) property: a section $\ell_\text{MRT}$ satisfies the MRT property if for all the points $\mbx \in \ell_\text{MRT}$ the mean return time $\tbar$ from $\mbx$ back to $\ell_\text{MRT}$, having completed one full rotation, is constant. 

First constructed in \cite{SchPik13} by means of an algorithmic numerical procedure, the MRT phase was later related to the solution of a boundary value problem in \cite{CaoLin20}, in which it was shown that the $\ell_\text{MRT}$ sections correspond to the level curves of a function $T(\mbx)$ satisfying the following PDE associated with a first-passage-time problem
\begin{equation}\label{eq:meanIsochrons}
	\LLd [T(\mbx)] = -1, 
\end{equation}
where $\LLd$ corresponds to the Kolmogorov backwards operator defined in \e{kolmogorov-backwards} \footnote{Unlike the stochastic asymptotic phase, the MRT phase requires the operator $\LLd$ to be strongly elliptic \cite{mclean2000strongly}.}. Imposing a boundary condition amounting to a jump by $\tbar$ across an arbitrary section transverse to the oscillation \footnote{The exact formulation of the jump boundary conditions is given by
\begin{equation}
\lim_{\varepsilon \to 0^+} T(K(\phi+2\pi-\varepsilon, \boldsymbol{\eta})) - T(K(\phi+\varepsilon, \boldsymbol{\eta})) = \tbar,
\end{equation}
with $K(\phi, \boldsymbol{\eta})$ refers to the parametrization defined in \eqref{eq:the-transformation}.}, the \textit{unique} solution of  \e{meanIsochrons}, up to an additive constant $T_0$, is a version of the so-called MRT function,
\begin{equation}\label{eq:mrt-definition}
	\Theta(\mbx)=(2\pi/ \tbar)(T_0-T(\mbx)).    
\end{equation}
Combining \eqref{eq:meanIsochrons} and \eqref{eq:mrt-definition}, the MRT phase $\Theta(\mbx)$ satisfies
\begin{equation}\label{eq:MRTphase}
	\mathcal{L}^\dagger[ \Theta(\mbx)] = \frac{2 \pi}{\tbar},
\end{equation}  
and the transformation of  $\mbX(t)$ in \e{SDE} to the MRT phase $\Theta$ obeys the stochastic differential equation
\begin{equation}
	\label{eq:mrt-Dyn}
 	  \diff \Theta(\mbX) = \frac{2 \pi}{\tbar} \diff t + \nabla \Theta(\mbX)^\intercal \mbg(\mbX)  \diff \mbW (t), 
\end{equation}
so its mean evolves in a way which is formally analogous to the dynamics for the deterministic phase (see \e{det-phase}).

\section{Self-Contained Phase Equation}\label{sec:sec-3}

We have introduced two different phase mappings: the asymptotic phase $\Psi(\mbx)$ and the MRT phase $\Theta(\mbx)$, which yield two different equations \eqref{eq:psiDyn} and \eqref{eq:mrt-Dyn}, respectively. However, neither of these equations is fully self-contained, as they both depend on $\mbX(t)$ \footnote{Similarly, the SDE for the ``variational phase'' derived in \cite{Mac23} is not self-contained.}.

Given an arbitrary phase mapping 
\begin{equation}\label{eq:gen-phase-map}
	\begin{aligned}
		\Phi : \mathcal{D} \subset \mathbb{R}^n & \rightarrow\mathbb{T}\\
		\mbx & \to \Phi(\mbx), 
  \end{aligned}
\end{equation}
so the evolution of the phase observable $\Phi(\mbX(t))$ follows 
\begin{equation}\label{eq:gen-phase-dyn}
\diff \Phi(\mbX) = \mathcal{L}^\dagger[\Phi(\mbX)] \diff t + \nabla \Phi(\mbX)^\intercal \mbg(\mbX)  \diff \mbW (t),
\end{equation}
we aim to derive a reduction procedure leading to a self-contained equation of the form 
\begin{equation}
	\label{eq:ideal-phase}
	\diff\phi = a_\phi(\phi) \diff t + \sqrt{2D_\phi(\phi)}\diff W_\phi(t),
\end{equation}
where $\diff W_\phi$ is the increment of a single Brownian motion (rather than $k$ of them) and $D_\phi$ is a (phase-dependent) effective noise intensity. 
Both $D_\phi$ and the phase-dependent  
local frequency $a_\phi$ should be smooth and periodic in $\phi$ (such that $a_\phi(\phi) = a_\phi(\phi+2\pi)$ and $D_\phi(\phi) = D_\phi(\phi+2\pi)$). While \e{ideal-phase} is a fully self-contained phase equation, it presents the challenge of estimating these new functions $a_\phi(\phi)$ and $D_\phi(\phi)$ in such a way the reduced and self-contained dynamics $\phi(t)$ in \eqref{eq:ideal-phase} approximate as precisely as possible the full phase dynamics $\Phi(\mbX(t))$ in \eqref{eq:gen-phase-dyn}. 

\subsection{Phase Mapping Requirements} \label{sec:requirements}

For our reduction procedure, we need the phase mapping $\Phi(\mbx)$ in \eqref{eq:gen-phase-map} to  satisfy  the following set  of conditions. We assume $\Phi(\mbx)$ to %unequivocally 
uniquely assign a single and well defined phase to each point $\mbx \in \mathcal{D}$ and to be continuous (at least $C^2$). 
This way, we assume it is possible to parameterize phase level sets (the \textit{isochrons}) by means of a set $\boldsymbol{\eta} = (\eta_1, \dots, \eta_{n-1})$ of $n-1$ amplitude-like variables, with $\eta_i(\mbx) \in \mathbb{R} \quad\forall i \in \{1, \dots, n-1\}$. 
This requirement amounts to assuming the existence of an \textit{invertible} parametrization $\mbx=K(\phi, \boldsymbol{\eta})$ 
	\begin{equation}\label{eq:the-transformation}
		\begin{aligned}
			K :  \mathbb{T} \times \mathbb{R}^{n-1} & \rightarrow \mathcal{D} \subset \mathbb{R}^n\\
			(\phi, \boldsymbol{\eta}) &\rightarrow K(\phi, \boldsymbol{\eta})
		\end{aligned}
	\end{equation}
	with $K(\phi, \boldsymbol{\eta}) = K(\phi+2\pi, \boldsymbol{\eta})$. We remark that since the phaseless sets (i.e, the points in which the phase function is not defined) are not invertible, neither they nor a small $\epsilon$-ball around them can belong to the domain $\mathcal{D}$.
 Thus the isochrons provide a foliation of the domain, $\mathcal{D}=\bigcup_{\phi\in[0,2\pi)}\mathcal{I}_\phi$, where the $\mathcal{I}_\phi$ are nonempty, simply connected, and pairwise disjoint.

\subsection{Reduction framework} \label{sec:reduction_framework}

Consider the general phase observable $\Phi(\mbx)$ in \e{gen-phase-map}, and assume the conditions given in \ref{sec:requirements} are satisfied. %Its evolution equation is derived by means of the Itô chain rule:
%\begin{equation}\label{eq:gen-phase-dyn}
%d\Phi(\mbx) = \mathcal{L}^\dagger[\Phi(\mbx)] \diff t + \nabla \Phi(\mbx)^\intercal \mbg(\mbx)  \diff \mbW (t).
%\end{equation}
Given that the sum of uncorrelated Gaussian white noise processes is Gaussian white noise, we rewrite \e{gen-phase-dyn} with one dimensional Gaussian white noise $\diff W_{\text{1D}}$,  such that:
\begin{equation}\label{eq:varthetaDyn}
    \diff \Phi(\mbX) = \mathcal{L}^\dagger[\Phi(\mbX)]\diff t + \sqrt{2D(\mbX)}\diff W_{\text{1D}},
\end{equation}
where the new noise amplitude term is given by
\[D(\mbx) = \frac{1}{2}\sum_{ijk} g_{ij}(\mbx)g_{kj}(\mbx)\partial_i\Phi(\mbx)\partial_k\Phi(\mbx). \]

%\bl{Do we need this sentence?}\rot{There exists an invertible transformation $\mbx=K(\phi, \boldsymbol{\eta})$ \eqref{eq:the-transformation}}. 
In what follows, we show a way in which the system in  \e{varthetaDyn} can be approximated by a reduced, self-contained equation of the form in  \e{ideal-phase}. In a nutshell, we rely on the existence of the invertible transformation $\mbx=K(\phi, \boldsymbol{\eta})$ in \eqref{eq:the-transformation} to integrate out the $n-1$ transverse directions, leaving only the phase dependency. For simplicity, we assume that we are in the planar case ($n=2$), so we only have one transverse variable, $\eta_1 = \eta$, to integrate out.

To start, we rewrite the stationary probability density $P_0(\mbx)$ in terms of this new set of coordinates $\mbx = K(\phi, \eta)$, and obtain the distribution $\bar{P}_0( \phi, \eta)$. We use it to define the following \textit{conditional} probability 
\begin{equation}\label{eq:cond-density-angular}
    \bar{P}_0(\eta | \phi) \equiv \frac{\bar{P}_0( \phi,\eta)}{\bar{P}_0(\phi)},
\end{equation}
provided the density $\bar{P}_0(\phi) > 0, ~\forall~ \phi\in\mathbb{T}$. Note that $K(\phi, {\eta}) = K(\phi+2\pi, {\eta})$, implying $\bar{P}_0(\eta | \phi) = \bar{P}_0(\eta | \phi + 2\pi)$.

Consider the dynamics for $\diff\Phi(\mbX)$ in  \e{varthetaDyn}. If we take the expected value of each side, using the stationary probability density, since $\diff W_{1D}(t)$ is independent of $\mbX(t)$ (and functions of $\mbX(t)$), we see that $\langle \sqrt{2D(\mbX(t))}\diff W_{1D} \rangle \equiv 0$. 
This motivates our choice of $a_\phi(\phi)$: we want $a_\phi(\phi)$ to represent the average rate of increase of $\Phi(\mbX(t))$ when $\mbX(t)$ happens to be on a particular isochron. 
That is,
\begin{equation}
    \label{eq:ideal-phase-reduction}   a_\phi(\phi) = \int_{\mbx \in \mathcal{I}_{\phi}} \mathcal{L}^\dagger[\Phi(\mbx(\phi,\eta))]\bar{P}_0(\eta|\phi)d\eta,
\end{equation}
where $\mathcal{L}^\dagger[\Phi(\mbx(\phi, \eta))]$ is the drift of $\Phi(\mbx)$ in \e{gen-phase-dyn}, which is averaged over the level curves of $\Phi(\mbx)$
\begin{equation*}
    \mathcal{I}_{\phi} = \{\mbx ~\in \mathcal{D}~|~{\Phi(\mbx)}={\phi}\},
\end{equation*}
which we parameterize by means of the transverse coordinate $\eta$. The $2\pi$-periodicity of both $K(\phi, \eta)$ and $\bar{P}_0(\eta | \phi)$ in $\phi$ implies $a_\phi(\phi) = a_\phi(\phi+2\pi)$.
%\rot{We note that, when computing the phase increments $\Delta \Phi$ in \eqref{eq:phase-increment} it is important to \textit{unwrap} the phase, that is to work with $\Phi(\mbx(t)) \in \mathbb{R}$ to avoid discontinuities (see SI for the complete details)}. \rot{However, when averaging, we need to bin the increments with respect to the \textit{wrap} phase. That is why we introduced the symbol $\wrapped{\phi}$

%
The choice of $a_\phi(\phi)$ in \e{ideal-phase-reduction} is meant to ensure that our reduction captures the first moment of the short term dynamics of $\Phi(\mbX(t))$. Assuming stationarity,  \e{ideal-phase-reduction} is equivalent to 
\begin{equation}\label{eq:ideal-phase-drift}
a_\phi(\phi) = \lim_{\diff t \to 0}~ \frac{1}{\diff t} \lr{\Delta \Phi(\mbX(t))}_{{\Phi}(\mbX(t)) ={\phi}}.
\end{equation} 
with
\begin{equation}\label{eq:phase-increment}
    \Delta \Phi(\mbX(t)) = \Phi_\text{unwrapped}(\mbX(t+\diff t)) - \Phi_\text{unwrapped}(\mbX(t)),
\end{equation} the increment of the phase variable between $t$ and $t+\diff t$. Computing phase increments requires omitting phase resets at $2\pi$. This is why we introduce $\Phi_\text{unwrapped}(\mbx)$, which maps $\mbx$ onto the real line $\mathbb{R}$ instead of the circle $\mathbb{T}$, such that 
\begin{equation*}
    \Phi(\mbx) = \Phi_\text{unwrapped}(\mbx) \mod{2\pi}.
\end{equation*} %\apc{We note that  averaging the phase increments \textit{on the circle} (i.e. $\lr{\cdot}_{\Phi(\mbx)}$ with $\Phi(\mbx) \in \mathbb{T}$) ensures the computation of $a_\phi(\phi)$ via \e{ideal-phase-drift} to also satisfy $a_\phi(\phi) = a_\phi(\phi+2\pi)$.} 
%(please note, $\Phi_\text{unwrapped}(\mbx) \mod{2\pi} = \Phi(\mbx)$).}
%\footnote{\rot{Note that computing phase increments requires taking the \textit{unwrapped} phase, that is, the phase projected onto the real line $\mathbb{R}$ instead of the circle, omitting the phase resets at $2\pi$, such that $\Phi_\text{unwrapped}(\mbx) \mod{2\pi} = \Phi(\mbx)$. This is to avoid jumps around $\phi = 0, 2\pi$.}}.\bl{Maybe here we can detail a bit.}
Once our choice of $a_\phi(\phi)$ is thus made, we choose $D_\phi(\phi)$ such that we best capture the second moment of the short-term dynamics of the phase:
\begin{equation}\label{eq:ideal-phase-diffusion}
    D_\phi(\phi) = \lim_{\diff t \to 0}~ \frac{1}{2\diff t}\lr{(\Delta \Phi(\mbX(t)) - a_\phi(\phi)\diff t)^2}_{{\Phi}(\mbX(t)) = {\phi}}.
\end{equation}
Expanding this formula yields
\begin{equation}\label{eq:ideal-phase-diffusion-formula}
%\begin{aligned}
    D_\phi(\phi) =~  \frac{1}{2}\sum_{ijk}\int_{\mbx \in \mathcal{I}_{\phi}} g_{ij}(\mbx)g_{kj}(\mbx)\partial_i \Phi(\mbx) \partial_j \Phi(\mbx) \bar{P}_0(\eta|\phi)d\eta%\\
    %+~ & \frac{1}{2}\Big(\int_{\mbx \in \mathcal{I}_{\phi}} \mathcal{L}^\dagger [\Phi(\mbx)]^2 \bar{P}_0(\eta|\phi)d\eta - a_\phi(\phi)^2\Big), 
%\end{aligned}
\end{equation}
where again $\mbx = K(\phi, \eta)$. Using the same argument as for $a_\phi$, we find $D_\phi(\phi) = D_\phi(\phi+2\pi)$.
%Note that in this reduction framework, the variability of single realizations around the mean comes from two distinct sources: not only from the average diffusion value \rot{(first term in the sum)}, but also from the variance of the drift along a given isochron \rot{(second term in the sum)}. 

Finally, we note that using equations \e{ideal-phase-drift} and \e{ideal-phase-diffusion} make it possible to extract $a_\phi$ and $D_\phi$ from a stationary time series $\Phi(\mbX(t))$. %\rot{which is how we compute the coefficients in the worked examples presented below}. 
They are obtained as the first two Kramers-Moyal coefficients of the trajectory $\phi(t)$. See \cite{FriPei97,TouSch11, RagKan01} for examples of how to extract them from stochastic trajectories, and the Supplementary Information (SI) for numerical details.

\subsection{Stochastic Asymptotic Phase Reduction}\label{sec:asymp-phase-red}

Let us now apply the general framework derived above to obtain a reduced evolution equation for the stochastic asymptotic phase, of the form
\begin{equation}
	\label{eq:ideal-phase-asymp}
	\diff \psi = a_\psi(\psi) \diff t + \sqrt{2D_\psi(\psi)}\diff W_\psi(t).
\end{equation}
From equation  \e{psiDyn}, we find that the drift term takes the form
\begin{equation*}
    \label{eq:drift-term}
        a_\psi(\psi) = \omega_1 - \int_{\mbx\in \mathcal{I}_\psi}  \bar{P}_0(\eta|\psi) \Omega(\mbx(\psi,\eta))d\eta.
\end{equation*}
Similarly, the effective noise term takes the form
\begin{equation}\label{eq:full-dif-asymp}
%\begin{aligned}
    D_\psi(\psi) = \frac{1}{2}\sum_{ijk}\int_{\mbx \in \mathcal{I}_{\psi}} g_{ij}(\mbx)g_{kj}(\mbx)\partial_i \Psi(\mbx) \partial_j \Psi(\mbx) \bar{P}_0(\eta|\psi)d\eta%\\
    %&+ \frac{1}{2}\Big(\int_{\mbx \in \mathcal{I}_{\psi}} (\omega_1 - \Omega(\mbx))^2 \bar{P}_0(\eta|\psi)d\eta - a_\psi(\psi)^2\Big),
%\end{aligned}
\end{equation}
where $\mbx=K(\psi, \eta)$ and $\mathcal{I}_\psi$ refers to
\begin{equation*}
    \mathcal{I}_\psi =  \{ \mbx ~|~ {\Psi}(\mbx) = {\psi} \}.
\end{equation*}
As noted previously, under the stationarity assumption, those two expressions can be approximated from time series realizations of $\Psi(\mbX(t))$ using expressions \e{ideal-phase-drift} and \e{ideal-phase-diffusion}.% \rot{(see SI for more details)}.
\subsection{Mean--Return-Time Phase Reduction}

Let us now derive a reduced phase equation for the Mean--Return-Time phase, of the form
\begin{equation}
	\label{eq:ideal-phase-mrt}
	\diff \theta = a_\theta(\theta) \diff t + \sqrt{2D_\theta(\theta)}\diff W_\theta(t).
\end{equation}
Following the same general reduction procedure, we obtain  expressions for the corresponding drift function
\begin{equation}
\begin{aligned}
    a_\theta(\theta) &= \frac{2 \pi}{\tbar},
\end{aligned}
\end{equation}
 and the effective noise function
\begin{equation}\label{eq:full-dif-mrt}
    D_\theta(\theta) = \! \sum_{i,j} \! \int_{\mbx\in \mathcal{I}_\theta} \!\!\!  \bar{P}_0(\eta|\theta) \partial_i \Theta(\mbx) \partial_j \Theta(\mbx)   \mathcal{G}_{ij}(\mbx)d\eta,
\end{equation}
where $\mbx=K(\theta, \eta)$, and $\mathcal{I}_\theta$ accounts for the MRT sections
\begin{equation*}
    \mathcal{I}_\theta =  \{ \mbx ~|~ {\Theta(\mbx)} = {\theta} \}.
\end{equation*}

\section{Numerical simulations}\label{sec:sec-4}

We apply our framework to two 2D systems exhibiting canonical bifurcations and illustrating various mechanisms of noisy oscillations. We use a finite-differences scheme to compute the stochastic asymptotic phase $\Psi(\mbx)$ and the MRT phase $\Theta(\mbx)$, as well as the stationary density $P_0(\mbx)$, and equations \e{ideal-phase-drift} and \e{ideal-phase-diffusion} to obtain the drift functions $a_\psi, a_\theta$ and the diffusion functions $D_\psi, D_\theta$. We refer the reader to the Supplementary Information (SI) for the complete numerical details.

\begin{figure}[t]
\centering
\includegraphics[width=0.48\textwidth]{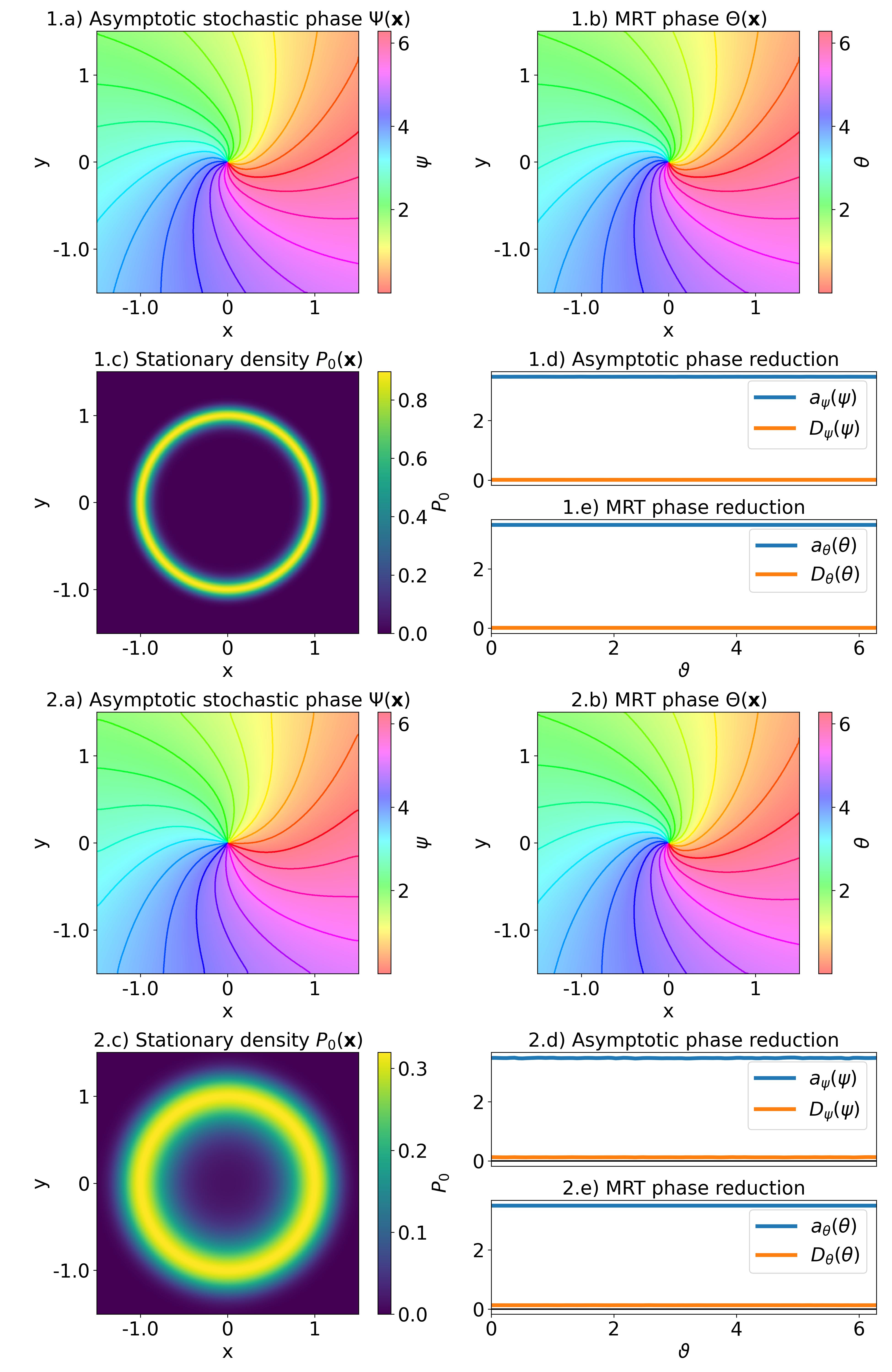}
\caption{\textbf{Hopf above bifurcation.} For the noisy Hopf bifurcation model in \e{slModel} with $\delta = 1$, $\beta= 0.5$, $\gamma = 4$, $\kappa = 1$, we show for two levels of noise (panel 1, above $D= 0.01$ and panel 2, below $D= 0.08$): (a) The asymptotic phase function $\Psi(\mbx)$. (b) The MRT phase function $\Theta(\mbx)$. (c) The stationary probability distribution. (d) Top panel shows the functions $a_\psi$ and $D_\psi$ and bottom panel shows $a_\theta$ and $D_\theta$, abscissa shared.}
\label{fig:Fig_1_Hopfa}
\end{figure} 
\begin{figure}[t]
\centering
\includegraphics[width=0.48\textwidth]{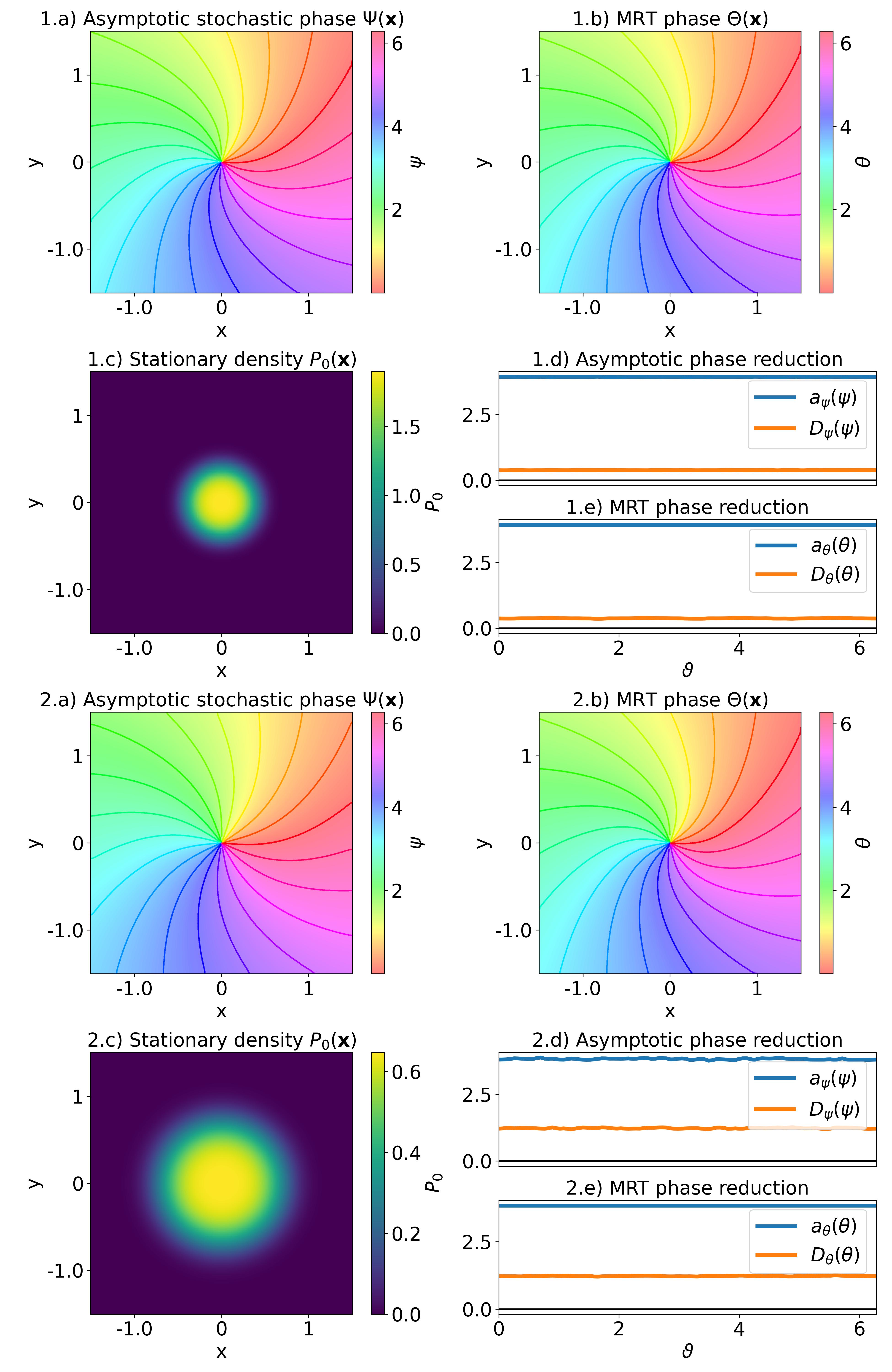}
\caption{\textbf{Hopf below bifurcation.} For the noisy Hopf bifurcation model in \e{slModel} with $\delta=-0.01$, $\beta= 0.5$, $\gamma = 4$, $\kappa = 1$, we show for two levels of noise (panel 1, above $D= 0.01$ and panel 2, below $D= 0.08$): (a) The asymptotic phase function $\Psi(\mbx)$. (b) The MRT phase function $\Theta(\mbx)$. (c) The stationary probability distribution. (d) Top panel shows the functions $a_\psi$ and $D_\psi$ and bottom panel shows $a_\theta$ and $D_\theta$, abscissa shared.}
\label{fig:Fig_1_Hopfb}
\end{figure}

\emph{Noisy Hopf bifurcation --} We consider the canonical model for a supercritical Hopf bifurcation endowed with Gaussian white noise
\begin{equation}\label{eq:slModel}		
\begin{aligned}
	\diff X &= \left[(\delta - \kappa R^2)X - (\gamma - \beta R^2)Y\right]\diff t + \sqrt{2D}\,\diff W_x(t),\\
	\diff Y &= \left[(\gamma - \beta R^2)X + (\delta - \kappa R^2) Y\right]\diff t +\sqrt{2D}\,\diff W_y(t),
\end{aligned}
\end{equation}
$R = \sqrt{X^2 + Y^2}$. In what follows, $\beta= 0.5$, $\gamma = 4$, $\kappa = 1$ \cite{TanChek2020}. In the deterministic setting, there is a supercritical Hopf bifurcation at $\delta = 0$. For $\delta>0$, there is a stable limit cycle $\Gamma$ of radius $R_* = \sqrt{\delta/\kappa}$ and period $T = \frac{2\pi}{\gamma - \beta R_*^2}$, which can be parameterized using the phase function $\vartheta(x,y) = \arctan(y/x) - \frac{\beta}{\kappa} \log\frac{r}{R_*}$ \cite{TanChek2020}. %$\sqrt{\beta}$ and period $T=2\pi$, which can be parameterized using the polar phase \cite{CasGui13}
%\[\theta = \arctan{\left(\frac{y}{x}\right)}.\]
The stochastic version has been studied for a long time, especially with respect to its correlation statistics (see e.g. \cite{HemLax67,UshWun05,JulDie09}), and, more recently, by means of the spectrum of $\LLd$ \cite{TanChek2020}.
%In Fig.~\ref{fig:Fig_1_Hopfa}, we show properties of the noisy system above the bifurcation ($\delta = 1$). We observe that for low noise ($D= 0.01$, panel~1) and high noise ($D= 0.08$, panel~2), the phase functions $\Psi(\mbx)$ and $\Theta(\mbx)$ still exhibit the characteristic \bl{rectilinear} structure \bl{modulated by the ``twist factor" $\beta/\kappa$} (see panels 1-2 (a) and 1-2 (b)) that would be present without noise \cite{TanChek2020}. Panels (1-2(c)) show the stationary probability densities. The trajectories are dispersed around the LC with radial symmetry dispersion which increases with the level of noise: because of the symmetry of both the model and the noise, the drift and effective diffusion terms are constant (see panels (d)).

\begin{figure}[t]
\centering
\includegraphics[width=0.48\textwidth]{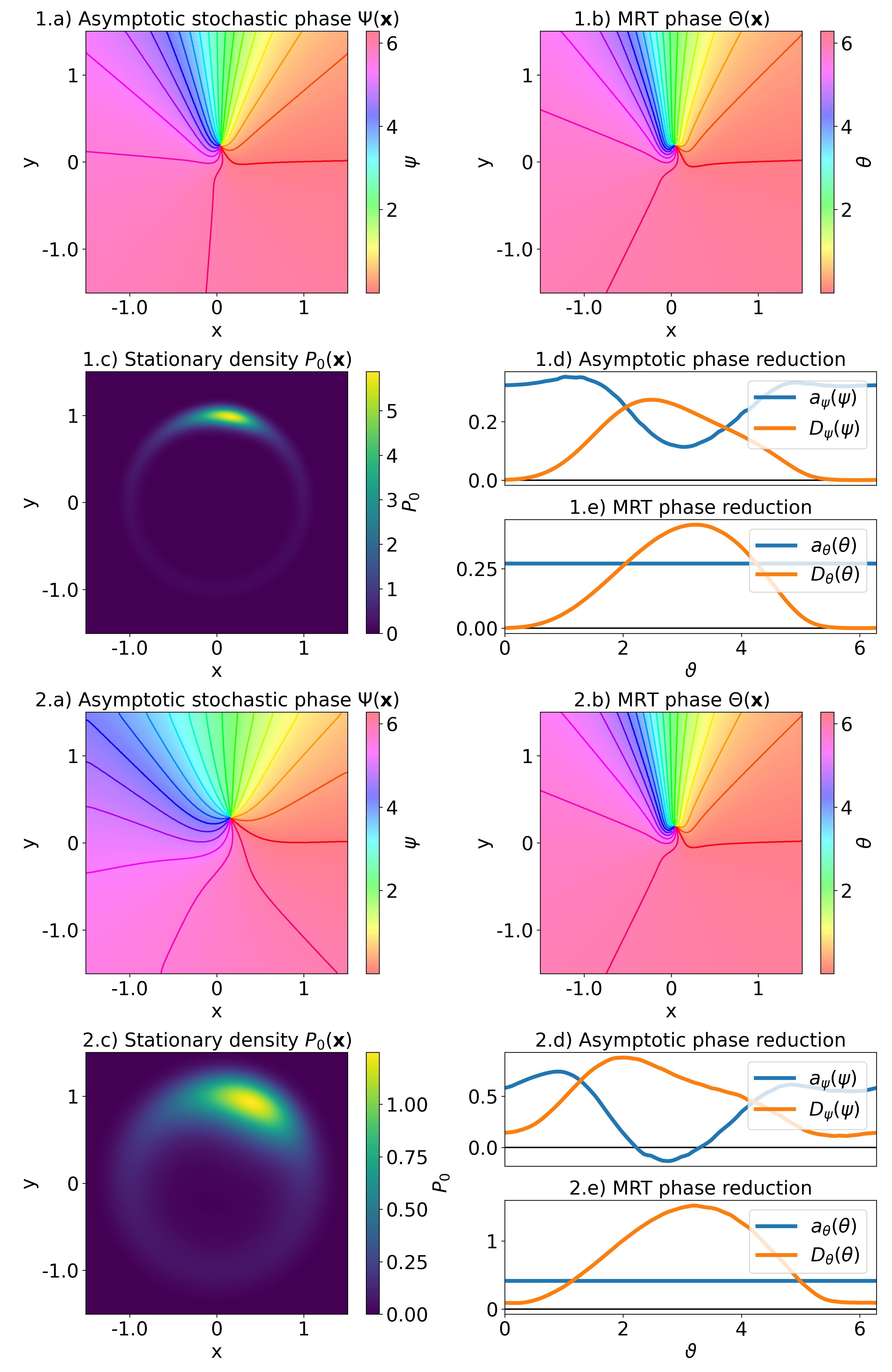}
\caption{\textbf{SNIC above bifurcation.} For the noisy SNIC bifurcation model in \e{snicModel} with $n = 1$, $m = 1.03$ we show for two levels of noise (panel 1, above $D= 0.01$ and panel 2, below $D= 0.08$): (a) The asymptotic phase function $\Psi(\mbx)$. (b) The MRT phase function $\Theta(\mbx)$. (c) The stationary probability distribution. (d) Top panel shows the functions $a_\psi$ and $D_\psi$ and bottom panel shows $a_\theta$ and $D_\theta$, abscissa shared.}
\label{fig:Fig_1_SNICa}
\end{figure} 
\begin{figure}[t]
\centering
\includegraphics[width=0.48\textwidth]{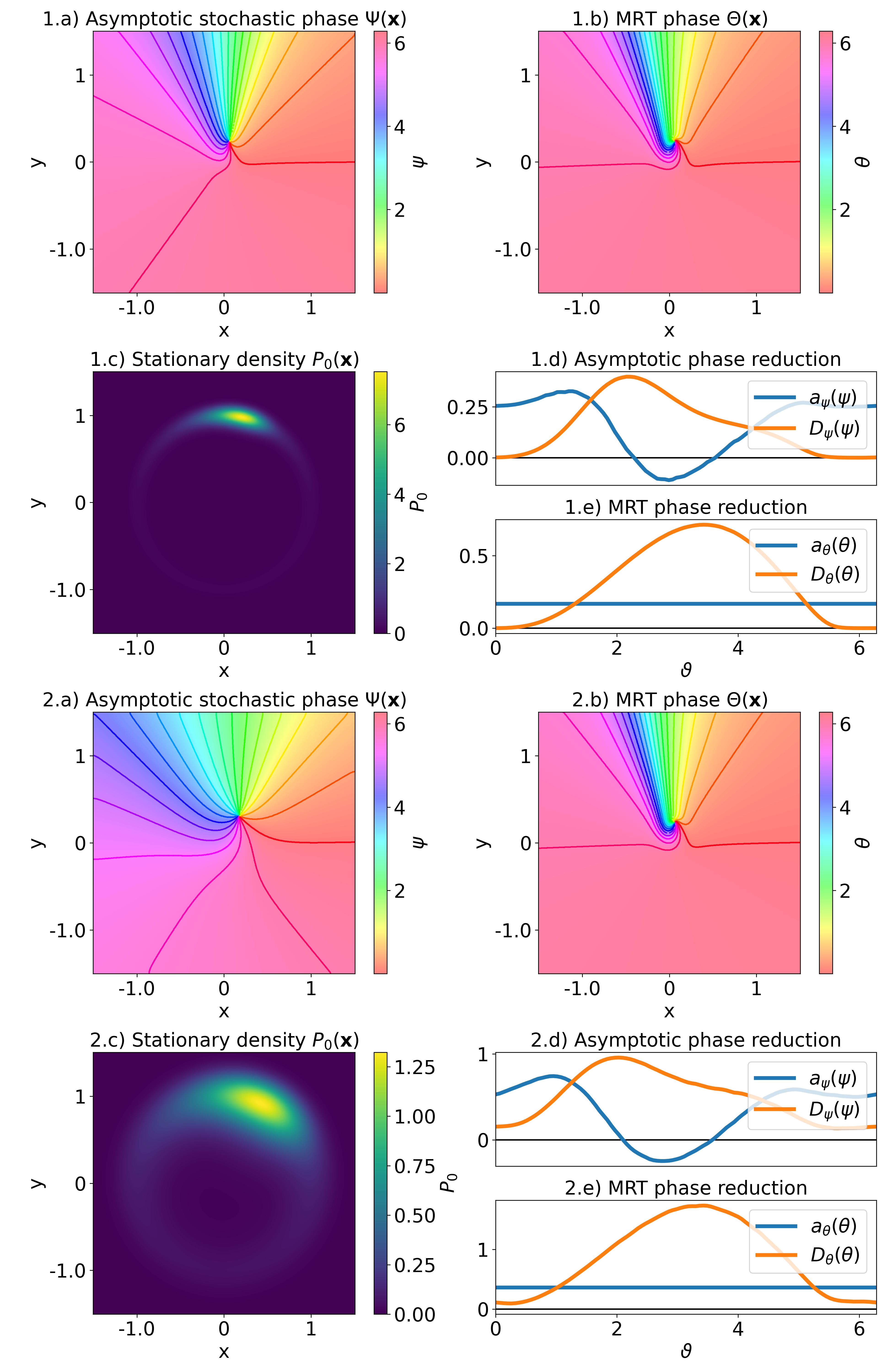}
\caption{\textbf{SNIC below bifurcation.} For the noisy SNIC bifurcation model in \e{snicModel} with $n = 1, m = 0.999$ we show for two levels of noise (panel 1, above $D= 0.01$ and panel 2, below $D= 0.08$): (a) The asymptotic phase function $\Psi(\mbx)$. (b) The MRT phase function $\Theta(\mbx)$. (c) The stationary probability distribution. (d) Top panel shows the functions $a_\psi$ and $D_\psi$ and bottom panel shows $a_\theta$ and $D_\theta$, abscissa shared.}
\label{fig:Fig_1_SNICb}
\end{figure} 

In Fig.~\ref{fig:Fig_1_Hopfa}, we show properties of the noisy system above the bifurcation ($\delta = 1$). We observe that, as we are considering additive isotropic noise, the phase functions $\Psi(\mbx)$ and $\Theta(\mbx)$ still exhibit the characteristic rotationally invariant rectilinear structure modulated by the ``twist factor" $\beta/\kappa$ that would be present without noise \cite{TanChek2020} (see panels 1-2 (a) and 1-2 (b)). Panels (1-2(c)) show the stationary probability densities for low and high noise amplitude($D = 0.01$ and $D = 0.08$, respectively). 
The trajectories are dispersed around the LC in a radially symmetric way, spreading increasingly with the level of noise. 
The constant drift and effective diffusion terms we recover (see panels (d)) reflect this rotational symmetry. 

Below the bifurcation, for $\delta < 0$, the deterministic system has a stable focus at the origin. 
Hence, in the absence of noise, the trajectories exhibit damped oscillations decaying towards the origin, and the asymptotic phase is not well defined \cite{ThoLin19}.
The addition of noise perturbs trajectories away from the stable steady state, leading to quasicycle oscillations \cite{BroBre15, PowLon21}. 
This is an example of noise-induced oscillations, leading to a non-zero probability of finding the system away from that fixed point. As can be seen in Fig.~\ref{fig:Fig_1_Hopfb}, panels c, the probability density has a 2D Gaussian-like profile, the maximum of which is located at the origin, where the deterministic fixed point is found. 
Despite the noise-induced character of the oscillation, the phase functions $\Psi(\mbx)$ and $\Theta(\mbx)$ have a structure similar to those of the LC case (panels (a) and (b)). 
As in the noisy LC case, the rotational symmetry of both $P_0$ and the phase mappings yields constant drift and effective noise functions (Fig.~\ref{fig:Fig_1_Hopfb}, panels 1-2 (d)).
However, we note that for the same levels of noise, the effective noise intensities $D_\psi$ and $D_\theta$ are much larger than in the LC case.

\emph{Saddle-node on an invariant circle --} Next, we consider a system that undergoes a saddle-node bifurcation on an invariant circle (SNIC) in the deterministic regime:
\begin{equation}\label{eq:snicModel}
	\begin{aligned}
	\diff X &\!\!=\!\!  \left[nX - mY - X R^2\! +\! \frac{Y^2}{R}\right]\diff t   +\sqrt{2D}\,\diff W_x(t),
	\\
	\diff Y &\!\!=\!\! \left[m X + n Y - Y R^2\! -\! \frac{XY}{R}\right]\diff t +\sqrt{2D}\,\diff W_y(t),
	\end{aligned}
\end{equation}
where $R(X,Y)=\sqrt{X^2+Y^2}$,
with $m, n \in \mathbb{R}$. We fix $n=1$, so in the noiseless case, there is an invariant curve $\Gamma$ of radius $\sqrt{n}=1$. When $m<1$, there are two fixed points onto $\Gamma$, a saddle and a node, which collide at $m = 1$, thus yielding an oscillatory LC state for $m>1$. The addition of noise to the saddle-node regime induces a non-zero probability that the system will leave the stable point and jump onto the circle, leading to oscillations. We refer to this state as the \textit{excitable} regime of the system. 

%Let us now present how our phase reduction applies to system \e{snicModel} in the oscillatory $m>1$ (Fig.~\ref{fig:Fig_1_SNICa}) and excitable cases $m<1$ (Fig.~\ref{fig:Fig_1_SNICb}). We consider parameter values close to the bifurcation ($m=1.03$ and $m=0.999$, respectively). In the low noise case (panels 1 in  Fig.~\ref{fig:Fig_1_SNICa} and  Fig.~\ref{fig:Fig_1_SNICb}), we observe that the phase functions we consider, that is $\psi(\mbx)$ (panel (a)) and $\Theta(\mbx)$ (panel (b)), have a similar structure reflecting the asymmetries in the velocity of the system during a cycle. Indeed, as the trajectories slow down near the ghost of the saddle-node, the phase sections appear more densely packed in this area of the phase space. By contrast, as trajectories speed up far from the ghost, we observe that the phase sections are more broadly separated. 

Let us now present how our phase reduction applies to system \e{snicModel} in the oscillatory ($m=1.03$) and excitable cases ($m=0.999$). In the low noise case (panel 1 in  Fig.~\ref{fig:Fig_1_SNICa} and  Fig.~\ref{fig:Fig_1_SNICb}, for the oscillatory and excitable regimes, respectively), we observe the phase functions $\Psi(\mbx)$ and $\Theta(\mbx)$ to have a similar structure reflecting the asymmetries in the velocity of the system during a cycle. 
As the trajectories slow down near the ghost of the saddle-node, the phase sections appear more densely packed in this area of the phase space. These velocity variations seem to be reflected in the drift term of the asymptotic phase reduction $a_\psi$: both above and below the bifurcation and for small and large noise values, we observe $a_\psi$ to be smaller (larger) near to (far from) the phase values corresponding to the locations of the saddle node (see panel 1-2 (d) in  Fig.~\ref{fig:Fig_1_SNICa} and  Fig.~\ref{fig:Fig_1_SNICb}).

\begin{figure}[t!]
\centering
\includegraphics[width=0.48\textwidth]{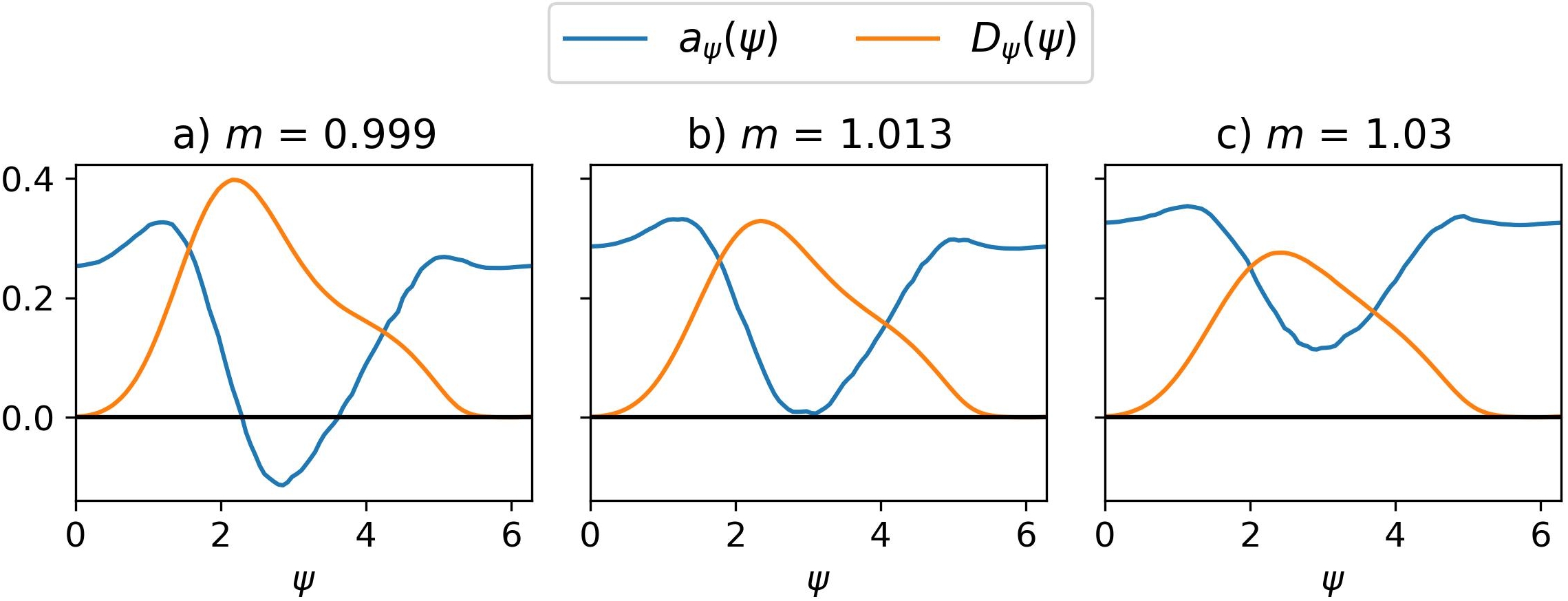}
\caption{\textbf{Transition across the SNIC bifurcation.} 
Asymptotic stochastic phase $\psi$ drift function $a_\psi(\psi)$ (blue) and effective diffusion term $D_\psi(\psi)$ (orange) for the SNIC for $n = 1$, $D=0.01$, and increasing values of $m$ across the bifurcation ($y$-axis shared). 
The transition from excitable to oscillatory regime appears to be reflected in the presence or absence of crossings in the drift function and in the decrease of the noise amplitude term. \textbf{Left}: $m = 0.999$. \textbf{Middle}: $m = 1.013$. \textbf{Right}: $m = 1.03$}
\label{fig:excitability}
\end{figure}

A closer look to our results reveals that for small levels of noise and below the bifurcation, the asymptotic phase drift $a_\psi$ shows two zero crossings (Fig.~\ref{fig:Fig_1_SNICb}.1-d). %This result for the excitable SNIC can be seen as a formulation of noise induced oscillations in excitable systems as a one dimensional escape problem. 
Indeed, there is a phase interval for which the drift term $a_\psi$ becomes negative: in that range, the phase $\psi$ behaves like a particle stuck in a one dimensional potential well and subjected to thermal fluctuations with noise intensity $D_\psi(\psi)$. 
A full rotation occurs when the fluctuations manage to push the phase out of the well.  Furthermore, we observe that, if we keep the noise weak and $m$ is varied so that the system goes above the bifurcation, $a_{\psi}$ becomes fully positive (Fig.~\ref{fig:Fig_1_SNICa}.1-d). We check in Fig.~\ref{fig:excitability} that this transition seems to be continuous, as the drift appears to smoothly move across the zero line as the bifurcation parameter $m$ is taken across $m=1$. By contrast, this transition across the zero line as the $m$ parameter is varied around $1$ does not hold anymore when noise levels get too large. Indeed, looking at Fig.~\ref{fig:Fig_1_SNICa} \& \ref{fig:Fig_1_SNICb}, panels 2, the drift terms $a_\psi$ show zero-crossings both below and above the bifurcation. In Appendix~\ref{app:SNIC-extra}, we provide additional details about the speed variations of the drift term $a_\psi$ and the emergence of zero crossings above the bifurcation as the noise is increased.

The interplay of deterministic and stochastic effects is apparent in the shape of the effective noise function $D_\psi$. 
As seen in panels 1-2 (d) in  Fig.~\ref{fig:Fig_1_SNICa} and  Fig.~\ref{fig:Fig_1_SNICb}, 
$D_\psi$ shows pronounced maxima near the phase values in which the mean velocity is minimal. 
This effect occurs both above and below the bifurcation, and is seen for both large and small noise levels.
The slowing down of the systems in these regions is consistent with the phase functions having more densely packed isochrons, which in turn corresponds to large phase gradients. 
Large values of both the gradients $\nabla \Psi$ and the stationary density $P_0$ near low-velocity areas explain the large values of $D_\psi$ in these regions.
These effects are also apparent in eq. \eqref{eq:full-dif-asymp}.
%We interpret this result as follows:  we are considering two cases that are very close in the parameter space, therefore, large levels of noise effectively blur the distinction between the two sides of the bifurcation. 

Finally, we comment the results of the reduction procedure applied to the MRT phase. 
By construction, the MRT phase defines sections with uniform mean return times. 
For this reason, one should not be surprised to find a constant drift term. 
In this case, all the variability in the velocity along the cycle is carried by the effective noise term $D_\Theta$. As for $D_\psi$, the collocation of high stationary density and large phase gradients leads to large values of the diffusion coefficient, cf.~\eqref{eq:full-dif-mrt}. We also see that the mean-return-time period $\tbar$ reflects the difference between regimes: $a_\theta = 2\pi/\tbar$ is smaller below bifurcation than above, meaning that $\tbar$ is larger below bifurcation than above.

\section{Checking the accuracy of the reduction via its long-term statistics}\label{sec:sec-5}
Our choice for the drift and effective noise coefficients of the reduced phase equation \eqref{eq:ideal-phase} ensures our reduction accurately captures the \textit{short term} ($\lim \diff t \to 0$) statistics of the full phase evolution. However, a meaningful phase reduction should also be able to reliably capture the \textit{long-term} (asymptotic) statistics of the evolution of the full system.  For that reason, given a general phase mapping as in \e{gen-phase-map}, we will consider the following statistics: the mean rotation rate
\begin{equation}\label{eq:long-term-drift}
\omega^\phi_{\text{eff}} = \lim_{t \to \infty} \frac{1}{t} \lr{{\Phi_\text{unwrapped}}(t) - {\Phi_\text{unwrapped}}(0)},
\end{equation}
and the phase diffusion coefficient
\begin{equation}\label{eq:long-term-diffusion}
D^\phi_{\text{eff}} = \lim_{t \to \infty} \frac{1}{2t} \lr{[{\Phi_\text{unwrapped}}(t) - {\Phi_\text{unwrapped}}(0) - \omega^\phi_{\text{eff}} t]^2},
\end{equation}
where we note that both statistics require using the \textit{unwrapped} phases as done in \eqref{eq:phase-increment}.
%where we take the \text{unwrapped} phases as done in \ref{sec:reduction_framework}.
We will numerically compute those two quantities from ensemble of realizations of: (i) the full phase mapping $\Phi(\mbX(t))$ (solution of \e{gen-phase-dyn}); and (ii) its corresponding self-contained phase reduction $\phi(t)$, solution of \e{ideal-phase}. We consider that the closer the values of both statistics for the full and the reduced system, the more accurately our reduction captures the full dynamics. %We consider a match to mean that our reduction accurately captures the full dynamics.

Additionally, since the general phase reduction \e{ideal-phase} is a 1D SDE with periodic drift and noise coefficients $a_\phi$ and $D_\phi$,
we can use the results in \cite{LinSch02}, which give us the following expressions for the mean rotation rate \e{long-term-drift}

\begin{equation}
\label{eq:benjamin-formula-rot}
    \omega^\phi_{\text{eff}} = \frac{2\pi(1 - e^{V(2\pi)})}{\int_0^{2\pi} I_+(\tilde{\phi})\diff\tilde{\phi}/\sqrt{D_\phi(\tilde{\phi})}},
\end{equation}
and the phase diffusion coefficient \e{long-term-diffusion}
\begin{equation}\label{eq:benjamin-formula-dif}
    D^\phi_{\text{eff}} = \frac{4\pi^2\int_0^{2\pi} I_{\text{-}}(\tilde{\phi})I_+^2(\tilde{\phi})\diff\tilde{\phi}/\sqrt{D_\phi(\tilde{\phi})}}{\left[\int_0^{2\pi} I_+(\tilde{\phi})\diff\tilde{\phi}/\sqrt{D_\phi(\tilde{\phi})}\right]^3},
\end{equation}
where
\[V(\phi) = -\int_0^{\phi}  \frac{a_\phi(\tilde{\phi})}{D_\phi(\tilde{\phi})}\diff\tilde{\phi}, \enskip \]
and
\[I_\pm(\phi)= \pm e^{\mp V(\phi)} \int_\phi^{\phi\pm 2\pi}  \frac{e^{\pm V(\tilde{\phi})}}{\sqrt{D_\phi(\tilde{\phi})}}\diff\tilde{\phi}.\]
The evaluation of the corresponding integrals is feasible as long as the noise intensity is not too small. Additional details regarding the numerical computation of equations \eqref{eq:long-term-drift} to \eqref{eq:benjamin-formula-dif} can be found in the SI.

\begin{figure}[t!]
\centering
\includegraphics[width=0.483\textwidth]{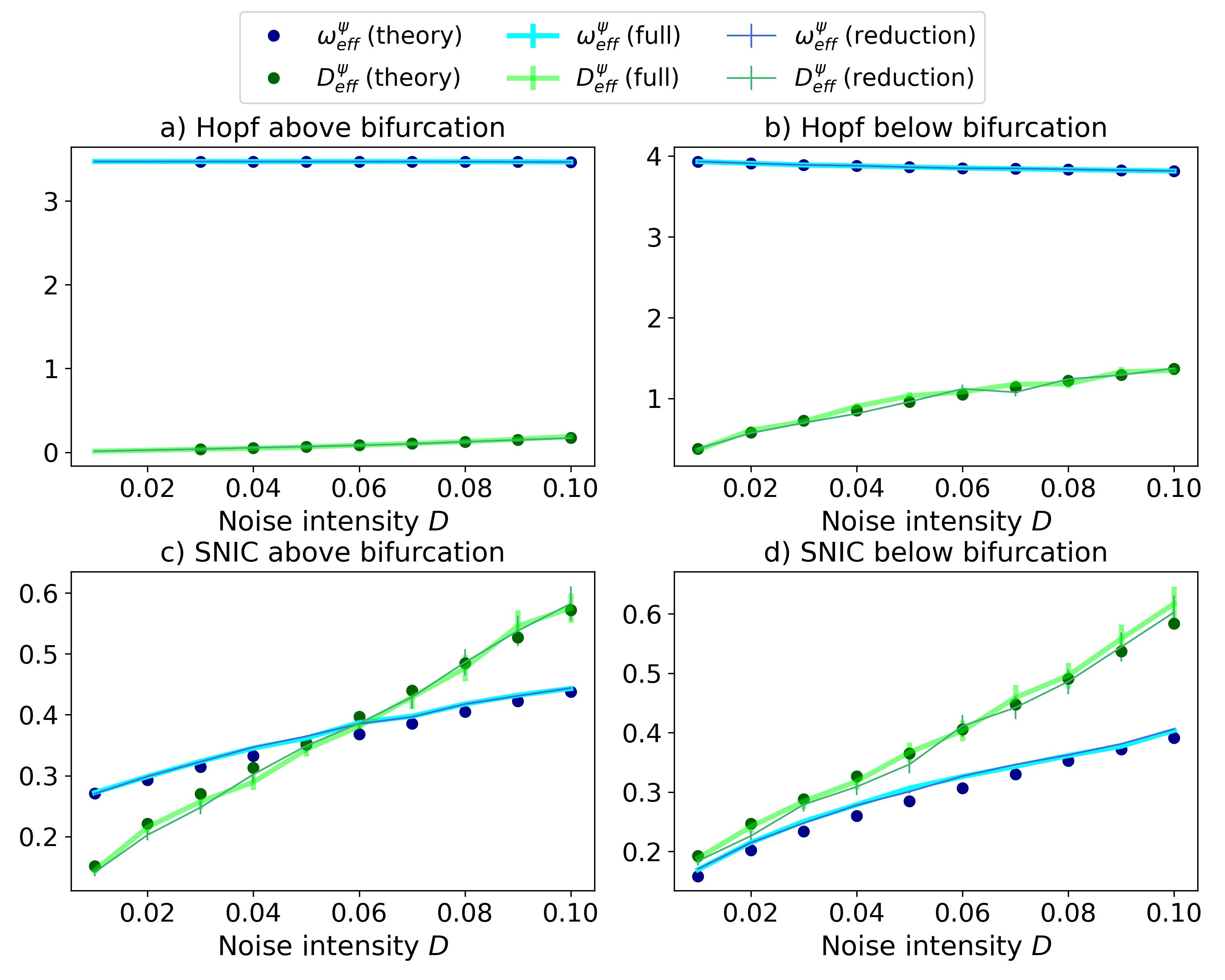}
\caption{\textbf{Long-term statistics of the asymptotic phase $\psi$ as a function of the noise strength $D$.} (a) Hopf bifurcation in the LC case; (b) Hopf bifurcation in the focus case; (c) SNIC in the LC case; (d) SNIC in the excitable case. We compute each statistic for: the full phase equation \e{psiDyn} (thick line), its phase reduction equation \e{ideal-phase-asymp} (narrow line) and the theoretical formulas \e{benjamin-formula-rot} and \e{benjamin-formula-dif} (dots).  Bars width of one standard error. Results for the MRT are similar and can be found in the Appendix~\ref{app:MRT-extra}}
\label{fig:stat_test}
\end{figure}

In Fig.~\ref{fig:stat_test} we show the long-term statistics of the full asymptotic phase $\Psi(\mbX(t))$ and of its corresponding reduction $\psi(t)$ as functions of the noise strength $D$. In general, we observe a good agreement between those quantities for both studied models. For the Hopf system, our phase reduction captures both the mean rotation rate $\omega^\psi_{\text{eff}}$ and the diffusion  coefficient $D^\psi_{\text{eff}}$ of the full system both in the LC and quasicycle regimes (panels (a) and (b)). We also observe a good agreement for the SNIC bifurcation both above and below the bifurcation (panels (c) and (d)). Appendix~\ref{app:MRT-extra} shows that repeating these calculations for the full MRT phase mapping $\Theta(\mbX(t))$ and its corresponding reduction $\theta(t)$ yields a similar level of agreement.

%\textit{Hopf bifurcation} -- As results in Fig.~\ref{fig:stat_test} panel (a) show, for the Hopf system in the LC case, our phase reduction captures both the mean rotation rate $\omega_{\text{eff}}$ and the diffusion  coefficient $D_{\text{eff}}$. Similarly, the reduction holds for the Hopf system below bifurcation, \rot{which is more surprising given that there is no underlying limit cycle}. \apc{Careful.}

%\textit{SNIC bifurcation} -- As our calculations for the long term statistics show (Fig.~\ref{fig:stat_test} panels c and d), our phase reduction accurately captures the mean rotation, both above and below the SNIC bifurcation. We only observe small deviations in $D_{\text{eff}}$ for high values of noise. We interpret this mismatch as a consequence of the pronounced fluctuations of the system's dynamics around the limit-cycle. When noise increases, our one-dimensional projection to a line cannot capture all details of the dynamics, as amplitude fluctuations play an important part in the system's dynamics. This \pt{approximation} leads to an underestimation of the long-term phase fluctuations. \apc{Careful. Now we have almost a match also in the SNIC case}

\section{Inferring Phase Response Properties}\label{sec:sec-6}

We will now show how our stochastic phase reduction framework can be applied to infer the phase response to weak external perturbations at linear order. In the case of a deterministic oscillator parameterized with phase $\vartheta$, the phase response to a weak perturbation can be linearized around the LC such that it is proportional to the gradient of the phase. This quantity is known as the \textit{infinitesimal phase response curve} (iPRC) \cite{ErmTer10}
\begin{equation}
\label{eq:iPRC}
\text{iPRC}(\vartheta) = \nabla \vartheta(\mbx)|_{\mbx = \gamma(\vartheta)},
\end{equation}
where $\mbx=\gamma(\vartheta)$ is the parametrization of the $\text{LC}$.

The main obstacle in finding a stochastic analogue of this quantity is the phase variability inherent to stochastic oscillators. As there is no LC, trajectories may visit any point $\mbx$ of the phase space with a given probability $P_0(\mbx)$. As a consequence, perturbing the system at the same phase will generally yield different phase responses. 

Consistent with our averaging approach to obtain a one-dimensional phase description, we postulate that, given a phase mapping $\Phi(\mbx)$ as the one defined in \eqref{eq:gen-phase-map}, a meaningful curve describing the mean response properties of the system can be obtained by averaging its gradient along a given isochron. We call this quantity the \textit{averaged iPRC (aiPRC)}, and can write it either in integral form
\begin{equation}\label{eq:aiPRC_P0}
\text{aiPRC}(\phi) = \int_{\mbx \in \mathcal{I}_{\phi}} \nabla \Phi(\mbx(\phi, \eta))\bar{P}_0(\eta|\phi)d\eta,
\end{equation}
or as an average across realizations
\begin{equation}\label{eq:aiPRC_trajectories}
\text{aiPRC}(\phi) = \lr{\nabla \Phi(\mbX(t))}_{{\Phi}(\mbX(t))={\phi}} .
\end{equation}
with both \eqref{eq:aiPRC_P0} and \eqref{eq:aiPRC_trajectories} satisfying $\text{aiPRC}(\phi) = \text{aiPRC}(\phi+2\pi)$. We remark that for systems with an underlying LC, in the vanishing noise limit ($D\rightarrow 0$), $P_0(\mbx) \rightarrow 0$ $\forall \mbx \notin \text{LC}$, and so \eqref{eq:aiPRC_P0} and \eqref{eq:aiPRC_trajectories} converge to the deterministic iPRC \eqref{eq:iPRC}.
\begin{figure}[t!]
\centering
\includegraphics[width=0.48\textwidth]{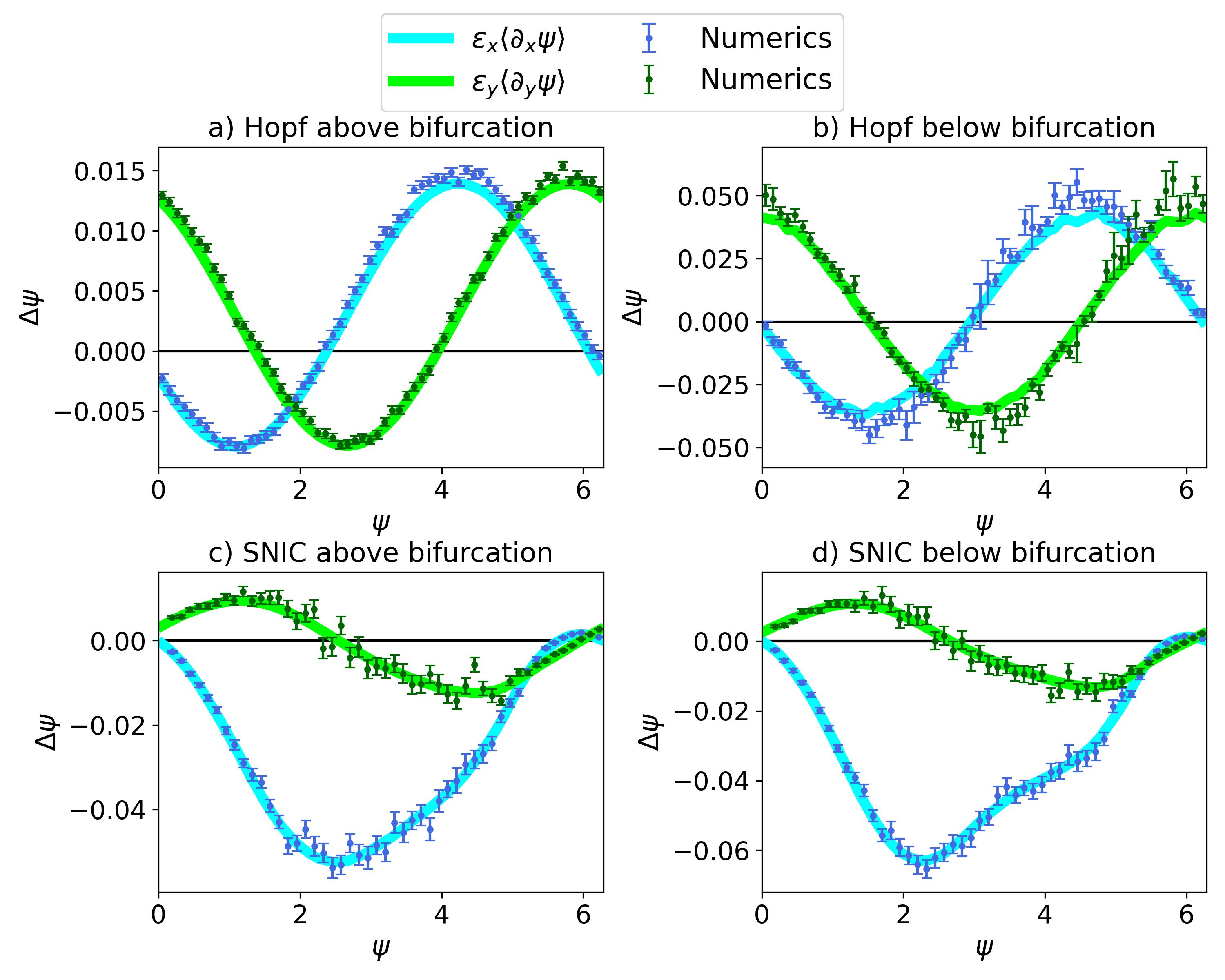}
\caption{\textbf{Averaged iPRCs for the asymptotic phase $\psi$}. Blue - response to a pulse in the X direction (amplitude $\epsilon_x$);  Green - response to a pulse in the Y direction ($\epsilon_y$) a) Hopf above bifurcation b) Hopf below bifurcation c) SNIC above bifurcation d) SNIC below bifurcation. External level of noise used for all systems: $D= 0.01$; pulse amplitudes: $\epsilon_x = \epsilon_y = 0.01$. Bars width of one standard error.}
\label{fig:sPRCs}
\end{figure}
We show that our aiPRC provides the expected phase shift $\Delta \Phi(\phi) = \phi_{\text{new}} - \phi$ of an oscillator subjected to a weak external pulse $\boldsymbol{\epsilon} \delta(t-t_0)$ as
\begin{equation}
    \Delta \Phi(\phi) \approx \boldsymbol{\epsilon} \cdot \text{aiPRC}(\phi).
\end{equation} 
In Fig.~\ref{fig:sPRCs}, we plot the aiPRC and compare it with numerical estimates of the average phase response, obtained by perturbing the system with a weak pulse at random phases, computing the individual phase shifts and binning the responses by phase. For each bin, we compute the average response using the circular mean \cite{MardJup00}. In panel (a) we compute the aiPRC for the Hopf normal form in the LC case (above the bifurcation). In this case, we observe that the aiPRC shows the characteristic sinusoidal Type II shape. Interestingly, a similar sinusoidal structure is found for the Hopf normal form for $\delta = -0.01$, when no LC exists. We note, however, that the amplitude of the mean response is larger in the quasicycle case than in the LC case. In this last case, the phase gradients dramatically increase near the origin where the probability density has a pronounced maximum. For the SNIC in the LC case (panel (c)), we observe that the aiPRC exhibits a Type I structure, very similar to the deterministic case (see \cite{GuiHug09} where this particular example is studied).  Interestingly, this structure is not much altered when the same object is studied below the bifurcation. We observe similar behavior for the aiPRC computed for the MRT phase (see Appendix~\ref{app:MRT-extra}).

\section{Extension beyond two  dimensions}\label{sec:extension-high-dim}

For the sake of illustration, the models we considered thus far were two-dimensional stochastic oscillators. 
We now discuss how our framework applies beyond the planar case ($n>2$). 
As discussed in sec.~\ref{sec:sec-3}, given a phase observable $\Phi(\mbx)$ as in \eqref{eq:ideal-phase}, its respective isochrons correspond to $n-1$ dimensional  manifolds which we assume can be parameterized by $n-1$ amplitude-like variables. 
For example, in the case $n=3$, the isochrons form 2D manifolds in the phase space, each of which can be parameterized by two amplitude-like variables (see \cite{MauMez18, PerMse20, Wilson2020} for examples of 3D isochrons and examples of such a parametrization for deterministic oscillators).

To show how our method extends to higher dimensional oscillators, we consider a stochastic version of the 3D Morris-Lecar neuron model, with slow delayed rectifier $K^+$ and subthreshold currents $I_{K,\text{dr}}$ and $I_{\text{sub}}$: 
\begin{align*}
    dV = &~ \frac1C\left[ I_{\text{ext}} - g_{\text{fast}}m_\infty(V)(V - E_{\text{Na}}) - g_{K,\text{dr}} Y (V - E_K) \right. \\ &\left. - g_{\text{sub}}Z(V - E_{\text{sub}}) - g_L(V - E_L)\right] \diff t + \sqrt{\frac{2D}{C}}\,\diff W(t)\\
    dY = &~ \phi_Y \frac{Y_\infty(V) - Y}{\tau_Y(V)}\diff t, \\ 
    dZ = &~ \phi_Z \frac{Z_\infty(V) - Z}{\tau_Z(V)}\diff t. \numberthis \label{eq:3DML}
\end{align*}
A variant of this model was
introduced in \cite{PreDeK2008}
using an Ornstein-Uhlenbeck process noise source as the stimulating current. 
Here, for simplicity, we adopt uncorrelated white noise in place of the OU process.
We choose the parameter values such that the neuron model exhibits class-I excitability, i.e. the system is poised at the edge of a SNIC bifurcation (see Appendix~\ref{app:ML-param} for the numerical values of parameters). The input-current fluctuations $\sqrt{2D}\xi(t)$ trigger noise-induced oscillations of the neuronal activity. This means that the duration between two spiking events, the interspike interval (ISI), is a random variable.

%and perform its stochastic phase reduction using the stochastic asymptotic phase $\psi$. Details and numerical values can be found in the Appendix.

Computing $\mathcal{L}^\dagger$ for higher dimensional systems is a nontrivial issue. 
To do so, we use generator Extended Dynamic Mode Decomposition (gEDMD), a recently developed data-driven method that relies on the knowledge of the underlying SDE to compute an approximation of $\mathcal{L}^\dagger$, its eigenvalues and its eigenfunctions, on an arbitrary basis of functions, in a judiciously chosen domain of the phase space \cite{KluNus20}. 
Using this approach allows us to 
perform our stochastic phase reduction using the stochastic asymptotic phase $\Psi(\mbx)$.
To lower the computational cost of calculating the eigenfunctions of $\LLd$ in higher dimensions, we use a long realization of the system to identify a region $\mathcal{R}$ of the phase space in which the oscillator typically resides (roughly corresponding to where the stationary density exceeds some small threshold), and compute an approximation of $\LLd$, its eigenfunctions, and thus of $\Psi(V, Y, Z)$, valid within this region. 
We then compute the self-contained phase reduction $\psi(t)$ in \eqref{eq:ideal-phase-asymp} by generating many trajectories of system \eqref{eq:3DML}.
Next we compute the drift $a_\psi$ and effective noise $D_\psi$ coefficients by averaging those many trajectories via the equations \eqref{eq:ideal-phase-drift} and \eqref{eq:ideal-phase-diffusion}. 
To evaluate the quality of the reduction, we compute the long-term statistics, as done in sec.~\ref{sec:sec-5}. 
We refer the reader to the SI for additional numerical details.

\begin{figure}[t!]
\centering
\includegraphics[width=0.483\textwidth]{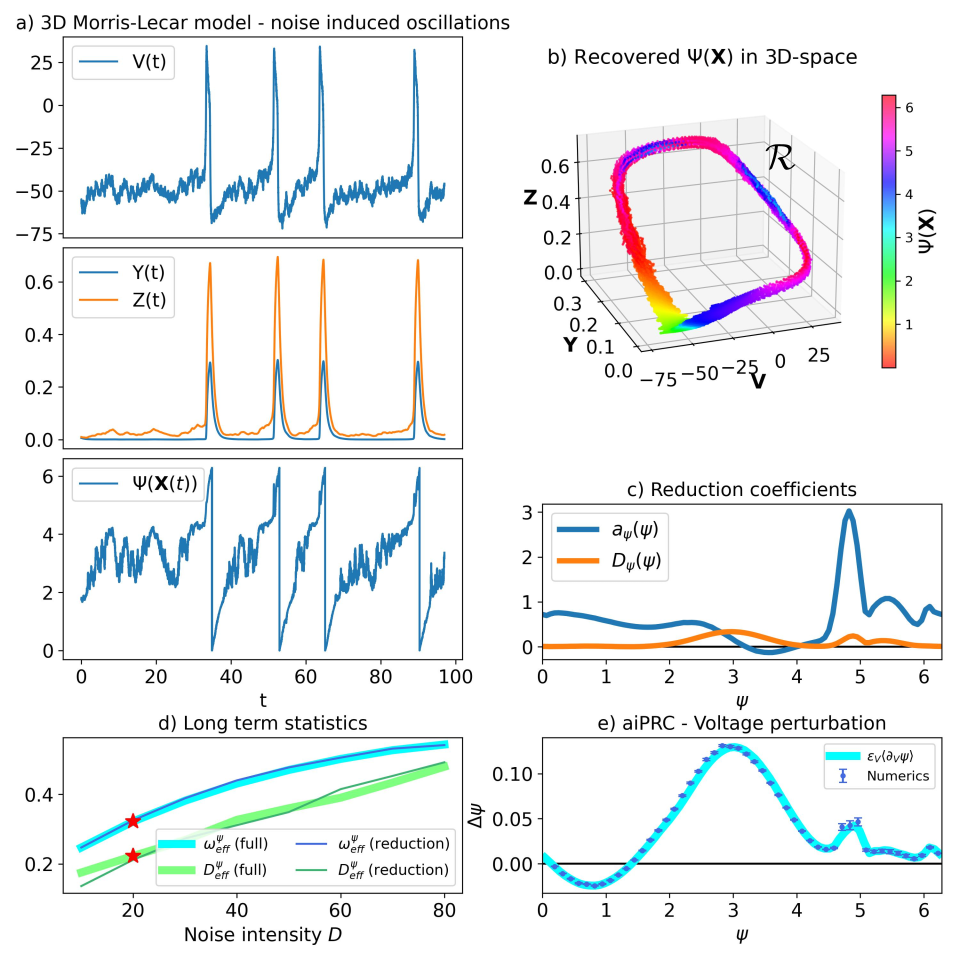}
\caption{\textbf{Asymptotic phase for the 3D Morris-Lecar model}. a) Typical realization of system \ref{eq:3DML} and recovered phase by means of gEDMD. 
b) Recovered asymptotic phase $\psi$ in the 3D phase space. 
Most realizations are confined around the region $\mathcal{R}$. c) Reduction coefficients d) Long-term statistics. e) Averaged iPRCs for the asymptotic phase for $\epsilon_V=1$. Noise intensity in panels a,b,c,e: $D=20$ (highlighted with a red star in panel d).}
\label{fig:ML}
\end{figure} 

We can further use gEDMD to compute the averaged iPRCs discussed in sec.~\ref{sec:sec-6}. Given that gEDMD approximates $Q_{\lambda_1}$ on  a judiciously chosen basis of functions $\{ F_i(\mbx) \}$ such that 
\begin{equation}
    \label{eq:Q-approx-as-a-finite-sum}
    Q^*_{\lambda_1}(\mbx) \approx \sum_i v_i F_i(\mbx),
\end{equation}
%$v_i \in \mathbb{C} \enskip \forall i$, % Why \forall i?
$v_i \in \mathbb{C}$, then we can also approximate its gradient as
\begin{equation}
    \label{eq:Q-grad-approx-as-a-finite-sum} 
    \nabla Q^*_{\lambda_1}(\mbx) \approx \sum_i v_i \nabla F_i(\mbx).
\end{equation}
In light of the identity
%\begin{align}\label{eq:gEDMD-PRC}
%    \nabla \Psi \equiv&~ \nabla \arctan\Big( \frac{\Im [Q^*_{\lambda_1}]}{\Re[Q^*_{\lambda_1}]} \Big) \\  =& \frac{\Re[Q^*_{\lambda_1}] \nabla \Im [Q^*_{\lambda_1}] - \Im [Q^*_{\lambda_1}] \nabla \Re[Q^*_{\lambda_1}]}{\Re[Q^*_{\lambda_1}]^2 + \Im [Q^*_{\lambda_1}]^2}.
%\end{align}
\begin{align}\label{eq:gEDMD-PRC}
    \nabla \Psi \equiv&~ \nabla \arctan\Big( \frac{\Im [Q^*_{\lambda_1}]}{\Re[Q^*_{\lambda_1}]} \Big) \\
    \nonumber
    =& \frac{\Re[Q^*_{\lambda_1}] \nabla \Im [Q^*_{\lambda_1}] - \Im [Q^*_{\lambda_1}] \nabla \Re[Q^*_{\lambda_1}]}{|Q^*_{\lambda_1}|^2},
\end{align}
the gradient of $\psi$ may be approximated by way of \eqref{eq:Q-approx-as-a-finite-sum}-\eqref{eq:Q-grad-approx-as-a-finite-sum}.
Then, using the expression for $\nabla \Psi(\mbx)$ in \eqref{eq:gEDMD-PRC} and assuming ergodicity, we can compute the aiPRC from time averages of realizations of the system using eq.~\eqref{eq:aiPRC_trajectories}. 

Fig.~\ref{fig:ML} displays the results of our gEDMD-based method.
Panel a) shows a typical realization of system \eqref{eq:3DML} and the corresponding stochastic asymptotic phase $\Psi(V(t), Y(t), Z(t))$. 
The phase captures both the oscillatory nature of the neuron's activity, and the fluctuations which lead to irregular oscillations. 
Panel
b) shows the asymptotic phase function $\Psi(\mbx)$
in the 3D phase space recovered via gEDMD. 
The phase increases steadily, on average, with traces of noise-induced backtracks. 
Panel c) shows the  coefficients $a_\psi$ and $D_\psi$ of the reduced dynamics.
As in the SNIC case, either in the LC regime with large noise levels or in the excitable regime, (cf.~Fig.~\ref{fig:Fig_1_SNICa} panel 2d and Fig.~\ref{fig:Fig_1_SNICb} panels 1-2d, respectively) $a_\psi$ shows a small region of negative mean drift.
In addition, for a narrow range of phase around $\psi\approx 4.5-5$, there is a pronounced peak in $a_\psi$, corresponding approximately to the triggering and upswing of the spike. 
In contrast, the effective noise term $D_\psi$ is maximal near $\psi\approx \pi$, corresponding to the regions of irregular voltage fluctuation during the plateau between spikes.  
Panel d) shows that the reduced phase equation captures the long-term dynamics of the full system  evolution, which we interpret as a sign that the reduced SDE is a satisfactory description of system \eqref{eq:3DML}. %from a statistical point of view. % What does it mean "a statistical point of view"?
Panel e) compares the expected phase response with the response obtained via direct perturbation. The responses match, showing that the procedure leading to the formula $\nabla\Psi$ derived from \eqref{eq:Q-approx-as-a-finite-sum}-\eqref{eq:gEDMD-PRC} allows to predict the response of the oscillator to an external pulse.

\section{Discussion}\label{sec:sec-discussion}

\textit{Summary.}
In this work, we developed a generalized self-contained stochastic phase reduction framework. 
Specifically, we provided a method for finding an approximate, self-contained phase reduction of stochastic oscillators subjected to Gaussian white noise. 
To illustrate our framework, we considered two mappings $\Phi:(\mbx)\in\R^n\to \mathbb{T}\equiv[0,2\pi)$, namely the ``Mean--return-time" phase $\Theta(\mbx)$ introduced in \cite{SchPik13} and the ``stochastic asymptotic phase" $\Psi(\mbx)$ introduced in \cite{ThoLin14}. 
%Even though our framework can be applied to $n$-dimensional systems, for clarity, we focused on examples with two-dimensional stochastic oscillators.
Throughout this work, we focused on examples of two-dimensional stochastic oscillators. However, the framework can be applied to $n$-dimensional systems: as a proof of concept, we applied our reduction procedure to a 3-dimensional neuron model.

Our reduction is built by considering the short term dynamics of the full phase variable. 
In order to test the accuracy of our reduction, we considered two well-known long-term statistics: the mean rotation $\omega^\phi_{\text{eff}}$, and the diffusion  coefficient $D^\phi_{\text{eff}}$. 
We consider the reduction to be better, the closer the agreement of these two statistics for i) the full phase system and ii) its reduced version. As previously mentioned, we have studied our reduction method for two different phase functions $\Theta(\mbx)$ and $\Psi(\mbx)$. 
Although the long-term statistics of a given oscillator are necessarily identical for both \textit{full} phase mappings, we have not found significant differences on the asymptotic statistics when considering a particular phase for reducing. Indeed, as we observe in Figs.~\ref{fig:stat_test}, \ref{fig:ML} and \ref{fig:MRT-aiPRC}, the long-term accuracy of both reductions, $\psi(t)$ and $\theta(t)$, remains good for all levels of noise in the range considered.

In contrast to the long-term behavior, we find differences between the considered phase mappings when it comes to the \textit{short-term} behavior. Indeed, we have found important differences in the drift functions, $a_\theta$ and $a_\psi$. 
The MRT phase has a constant drift term, so the phase dependence of its dynamics resides entirely in the effective noise term $\sqrt{2D_\theta(\theta)}$. 
By contrast, the stochastic asymptotic phase, shows a non-constant drift term. 
As we have shown in Fig.~\ref{fig:excitability}, this variable drift term, at small noise, reflects important \textit{dynamical} information about the system, namely the transition from an excitable to an oscillatory regime. 
Under strong noise conditions, however, zeros in the drift function can be observed both above and below the deterministic bifurcation point.

A way to understand this difference between drift functions is by relating our approach to the noise-induced frequency shift (NIFS) phenomenon \cite{YosAra08}:  
In a deterministic LC oscillator with phase $\vartheta$, adding white noise  typically causes a shift in the average frequency \cite{NewSch14}. 
The new average frequency is given by the ensemble average,
%\begin{equation}
    $\overline{\omega} = \langle \dot{\vartheta} \rangle.$
%\end{equation}
For a general stochastic oscillator, we see that, by construction, the MRT phase takes the effect of noise on the frequency into account by setting its instantaneous frequency to the average frequency: $a_\theta(\theta) = \overline{\omega} = \frac{2\pi}{\Tbar},$ for all $\theta$. 
By contrast, the asymptotic phase has an additional degree of variability, as it has an instantaneous average frequency value $a_\psi(\psi)$ which need not equal $\overline{\omega}$.
Thus, the asymptotic phase keeps track of finer details arising from the interaction between noise and deterministic dynamics, at the cost of added complexity in the equation.

\textit{Future Directions.} 
In this  paper, we have applied our method to systems whose SDE was known. 
However, both the MRT phase and the stochastic asymptotic phase can be extracted from data. 
For example, the original procedure to extract the MRT was built upon an iterative method that can accommodate both simulated and real-world data \cite{SchPik13}. 
The stochastic asymptotic phase was first extracted from data by fitting the oscillatory-exponential asymptotic decay of the probability density to its stationary state \cite{ThoLin14}. 
 The family of Dynamic Mode Decomposition (DMD) methods, such as gEDMD, based on an eigenfunction decomposition of the Koopman operator (which is closely related to the Kolmogorov backwards operator for stochastic systems) offer an alternative approach to obtaining these functions (see \cite{KluNus20}). 
Thus, as we have shown in sec.~\ref{sec:extension-high-dim} by means of gEDMD, they can allow one to recover an estimation of the spectral properties of $\mathcal{L}^\dagger$ from data, most particularly of the $Q^*_{\lambda_1}$ eigenfunction that carries the stochastic asymptotic phase, cf.~\cite{melland2023attractor}. 
This connection would allow application of our framework to real world oscillatory data, to be explored in future work.

In the deterministic case, adding an amplitude variable can extend the domain of accuracy of the phase description \cite{CasGui13, MonWil19, MauMez20, PerMse20}.
We believe our construction may benefit from a similar approach. 
Recently, the spectral analysis of $\LLd$ has been extended to provide an analogue of the so-called amplitude coordinates \cite{PerLin21, KatZhu21}. 
In related work, it has been shown that a different observable, the slowest decaying complex eigenfunction $Q^*_{\lambda_1}$ of the Kolmogorov backwards operator, yields a universal description of stochastic oscillators \cite{PerGut23}. 
This complex phase function, $Q^*_{\lambda_1}$, allows comparison of stochastic oscillators
regardless of their underlying oscillatory mechanism \cite{PerGut23}.
Written in polar form, the complex phase function $Q^*_{\lambda_1}=ue^{i\psi}$ defines both a notion of phase $\psi\in[0,2\pi)$ and an amplitude $u$ that captures the concentration or coherence of an oscillator's probability density. 
Both the stochastic analogues of the phase-amplitude description and the complex phase ideas appear as interesting targets for future research in the field of stochastic dynamics \cite{Gar04,Ris96,mezic2005spectral,CrnMac20,chekroun2020ruelle, HumAsh23}.

An additional interesting question arising from this work is the exploration of the \emph{averaged infinitesimal phase response curve (aiPRC)} function defined in sec.~\ref{sec:sec-6}. 
We have shown that it provides a meaningful estimation of the average phase response of a stochastic oscillator to a small pulselike perturbation. 
Being able to compute the average response of stochastic oscillators to external perturbations by means of the aiPRC is a first step towards the analysis of complex noisy oscillatory phenomena, such as synchronization among oscillators connected on networks \cite{HopIzh97, CabCas22,NicAll24}. 
In the past, defining those phenomena, such as noisy phase and frequency synchronization \cite{FreSch03}, or noise-enhanced phase-locking \cite{NeiSil98}, required using a deterministic notion of phase, such as the Hilbert phase, and extending it to the noisy case. The work we put forward in this manuscript builds upon recent notions of stochastic phase \cite{ThoLin14, SchPik13}. Thus, obtaining a reduction for those stochastic phases will allow to revisit those earlier results in a purely stochastic setting.
Moreover, Adams and MacLaurin have recently proposed a formal approach to deriving a self-contained stochastic differential equation for what they term the ``isochronal phase", for systems that have a particular invariant manifold structure (such as system with an underlying LC) see \cite{adams2025isochronal}.  
Application of their methods, drawn from rigorous analysis of stochastic partial differential equations, to the examples we present here, is an interesting opportunity for future investigation.\\

%\bl{Update citation for Adams and McLaurin [50]}

% The \nocite command causes all entries in a bibliography to be printed out
% whether or not they are actually referenced in the text. This is appropriate
% for the sample file to show the different styles of references, but authors
% most likely will not want to use it. \nocite{*}

\section*{Code availability}

All code used to produce the results shown in this work is available at \url{https://github.com/PHouzel/stocha-phase-red}. The gEDMD code was made publicly available by the authors of \cite{KluNus20} at \url{https://github.com/sklus/d3s/}.

\section*{Acknowledgments}
This research was funded by Agence Nationale pour la Recherche (ANR-17-EURE-0017, ANR-10IDEX-0001-02), ENS, CNRS and INSERM. A CC-BY public copyright license has been applied by the authors to the present document and will be applied to all subsequent versions up to the Author Accepted Manuscript arising from this submission, in accordance with the grant’s open access conditions. This work was supported in part by (i) NSF grant DMS-2052109, (ii) the Oberlin College Department of Mathematics, and (iii) by the National Science Foundation under Grant DMS-1929284 while the author was in residence at the Institute for Computational and Experimental Research in Mathematics in Providence, RI, during the ``Math + Neuroscience: Strengthening the Interplay Between Theory and Mathematics" program. PH acknowledges support from École Doctorale Frontières de l’Innovation en Recherche et Éducation. The author APC is a Serra Húnter Fellow and acknowledges support from Spanish Ministry of Science and Innovation grants (Projects No. PID2021-124047NB-I00 and PID-2021-122954NB-100).

\appendix
\section{Examples of Eigenvalue Spectra for Robustly Oscillatory Systems}
\label{app:spectra}

\begin{figure}[t!]
	\centering
	\includegraphics[width=0.483\textwidth]{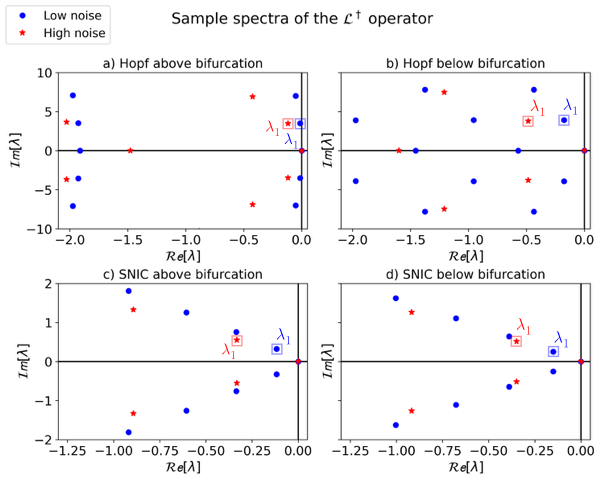}
	\caption{\textbf{Spectra of the backwards operator $\LLd$ for the planar models}. We plot the spectra of $\LLd$ for external noise amplitudes $D= 0.01$ (blue dots) and $D= 0.08$ (red stars). 
		The box indicates the non-trivial eigenvalue $\lambda_1$ with least real part and positive imaginary part. We indicate the values of $\lambda_1$ for weak (large) noise, respectively. Panel (a): $\lambda_1 = -0.01 + i3.5$, ($\lambda_1 = -0.12 + i3.48$). Panel (b): $\lambda_1 = -0.18 + i3.92 $, ($\lambda_1 = -0.48 + i3.78$). Panel (c): $\lambda_1 = -0.12 + i0.32$, ($\lambda_1 = -0.33+i0.55$). Panel (d): $\lambda_1 = -0.15+i0.25$, ($\lambda_1 = -0.35+i0.51$).}
	\label{fig:spectra}
\end{figure}

In this Appendix we show the spectra of the Hopf and SNIC models in the main text for the considered levels of noise $D=0.01$ and $D=0.08$. 
As the Fig.~\ref{fig:spectra} shows, in all the cases there exists a nontrivial eigenvalue of $\mathcal{L}^\dagger$ with least negative real part $\lambda_1 = \mu_1 + i\omega_1$, which is complex valued ($\omega_1>0$) and unique. Hence, as pointed out in \cite{ThoLin14} and explained in sec.~\ref{sec:asymp-phase}, the slowest decaying mode associated with $\lambda_1$ is complex, so one can extract the stochastic asymptotic phase $\Psi(\mbx)$ from the backward eigenfunction $Q_{\lambda_1}(\mbx)$. We also note that, consistent with the observations of \cite{TanChek2020}, in the Hopf case we observe a qualitative change in the shape of the principal eigenvalue family, which lies  approximately along a parabola above the bifurcation, and approximately along a straight line, as part of a checkerboard-like grid, below the bifurcation.

\section{Illustrating the SNIC case}
\label{app:SNIC-extra}

We provide additional details regarding the shapes of the drift term $a_\psi$ for the SNIC model that we observe in Fig.~\ref{fig:Fig_1_SNICa} and Fig.~\ref{fig:Fig_1_SNICb}. As discussed in sec.~\ref{sec:asymp-phase-red}, the evolution law of $\diff \Psi(\mbX(t))$ in \eqref{eq:psiDyn} yields a drift term $a_\psi(\psi)$ of the form
\begin{equation}\label{eq:psi-drift-app}
	a_\psi(\psi) = \omega_1 - \int_{\mbx\in \mathcal{I}_\psi}  \bar{P}_0(\eta|\psi) \Omega(\mbx)d\eta,
\end{equation}
with $\Omega(\mbx)$ defined in \eqref{eq:psiDyn}. To study the values for $a_\psi$ in the SNIC case, we define the following function
\begin{equation}\label{eq:psi-cal-drift-app}
	\mathcal{A}(\mbx) = P_0(\mbx) \left( \omega_1 - \Omega(\mbx) \right),
\end{equation}
where we weight the term $\omega_1 - \Omega(\mbx)$ by the stationary density $P_0(\mbx)$, making explicit the contribution of each point $\mbx\in\mathcal{D}$ when computing the average in \e{psi-drift-app}.

\begin{figure}[h!]
\centering
\includegraphics[width=0.483\textwidth]{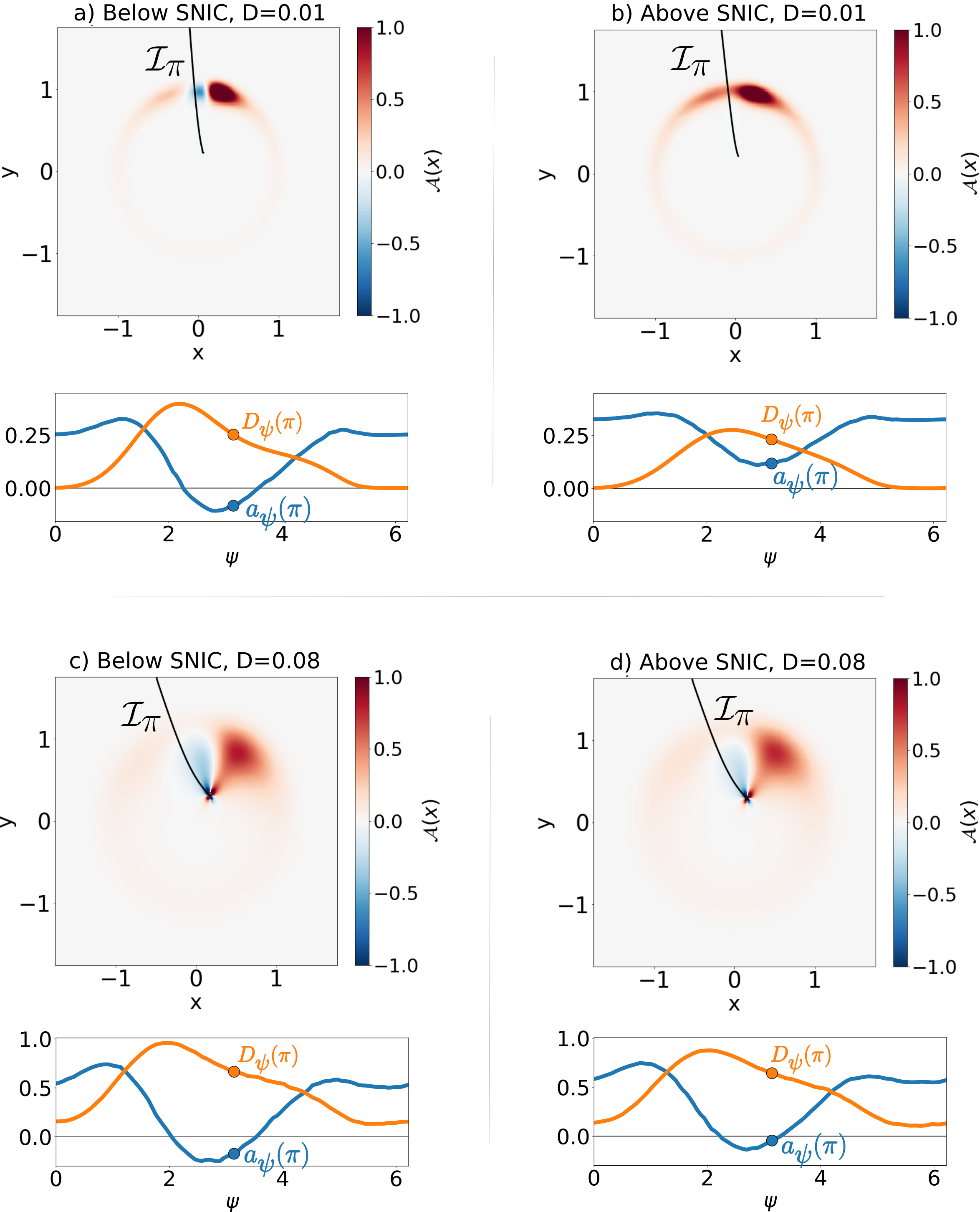}
\caption{\textbf{SNIC drift speed variations}. For small ($D= 0.01$) and large ($D= 0.08$) levels of noise, we plot the function $\mathcal{A}(\mbx)$ in \eqref{eq:psi-cal-drift-app} both below ($m=0.999$) and above ($m=1.03$) the SNIC bifurcation in system \eqref{eq:snicModel}. We also plot the corresponding drift $a_\psi(\psi)$ (blue) and diffusion $D_\psi(\psi)$ (orange) functions already shown in Fig.~\ref{fig:Fig_1_SNICa} and Fig.~\ref{fig:Fig_1_SNICb}. In black, we show the asymptotic phase isochron $\mathcal{I}_\pi$ corresponding to the phase $\psi=\pi$ and highlight the corresponding drift and diffusion terms recovered when averaging along $\mathcal{I}_\pi$.}
\label{fig:snic-appendix}
\end{figure} 

We summarize our results in Fig.~\ref{fig:snic-appendix}, in which we plot $\mathcal{A}(\mbx)$ for the SNIC model. For the sake of clarity, we also include the corresponding drift and diffusion terms $a_\psi$ and $D_\psi$ shown in Fig.~\ref{fig:Fig_1_SNICa} and Fig.~\ref{fig:Fig_1_SNICb} in the main text. 

In the small noise case ($D=0.01$), below the bifurcation (panel (a)), we observe the function $\mathcal{A}(\mbx)$ to be mainly positive with the exception of a tiny area in which $\mathcal{A}(\mbx)<0$. This area is the one to which we find the isochrons $\mathcal{I}_\psi$ such that $a_\psi(\psi) < 0$ (as is the case for e.g. with $\psi=\pi$). By contrast, above the bifurcation, we find $\mathcal{A}(\mbx) \geq 0~\forall\mbx\in\mathcal{D}$, such that $a_\psi(\psi) > 0$ $\forall\psi \in \mathbb{T}$. %As we have observed in the main text Fig.~\ref{fig:excitability}, the smooth transition from $a_\psi$ having negative values to be fully positive as the bifurcation parameter is varied around $m=1$, aligns very well with the observed transition from an excitable noise-induced regime to a noise-drive LC regime.

%$a_\psi(\pi) <0$ (see Fig.~\ref{fig:snic-appendix} panel 1a) and $\mathcal{A}$ clarifies why: the isochron $\mathcal{I}_\pi$ along which we average to obtain $a_\psi(\pi)$ visits points $\mbx$ in which $\mathcal{A}(\mbx)\leq0$. By contrast, above the bifurcation the result $a_\psi(\pi) >0$ found in the main text (shown also Fig.~\ref{fig:snic-appendix} panel 1b) concurs with the fact that $\mathcal{A}(\mbx) \geq 0$ for $\mbx\in\mathcal{I}_\pi$ (see panel (b) in Fig.~\ref{fig:snic-appendix}). It is also interesting to remark the decrease of the noise term as we move from the excitable to the oscillatory regime, thus reflecting the effect of the noise in the dynamics.  

By contrast, as we increase the external level of noise, we lose this smooth transition from a $a_\psi$ with negative values to a fully positive $a_\psi$: for $D=0.08$, the drift function displays a range of phases in which $a_\psi$ is negative both above and below the bifurcation. The emergence of ``noise-induced'' zeros in the drift term above the bifurcation as the noise increases can be illustrated by means of $\mathcal{A}(\mbx)$. Indeed, for $D=0.08$, we find $\mathcal{A}(\mbx)$ to be very similar both above and below the bifurcation (see panels (c) and (d) in Fig.~\ref{fig:snic-appendix}). %However, a closer comparison between (c) and (d) shows the range of phases in which the phase is negative to be larger below the bifurcation. 

%All together, we observe in all four panels in Fig.~\ref{fig:snic-appendix} that $\min(a_\psi)$ and $\max(D_\psi)$ are smaller and larger, respectively, below than above the bifurcation. We interpret this result as if the stochastic phase reduction was capturing the major role of the noise in the generation of oscillations when being below than above the bifurcation.

%Nevertheless, there is an important feature to observe: comparing panels (b) and (d) in Fig.~\ref{fig:snic-appendix}, we see that, while for small levels of noise above the bifurcation $\mathcal{A}(\mbx)\geq0 \enskip \forall~\mbx \in \mathcal{D}$, increasing the noise causes the appearance of areas where $\mathcal{A}(\mbx)<0$. Again, plotting the section $I_\pi$ clarifies why for the large noise case $a_\psi(\pi) < 0$ both above and below the bifurcation.

\section{MRT Extra Results: Long-Term Statistics \& aiPRCs}
\label{app:MRT-extra}

\begin{figure}[h]
\centering
\includegraphics[width=0.48\textwidth]{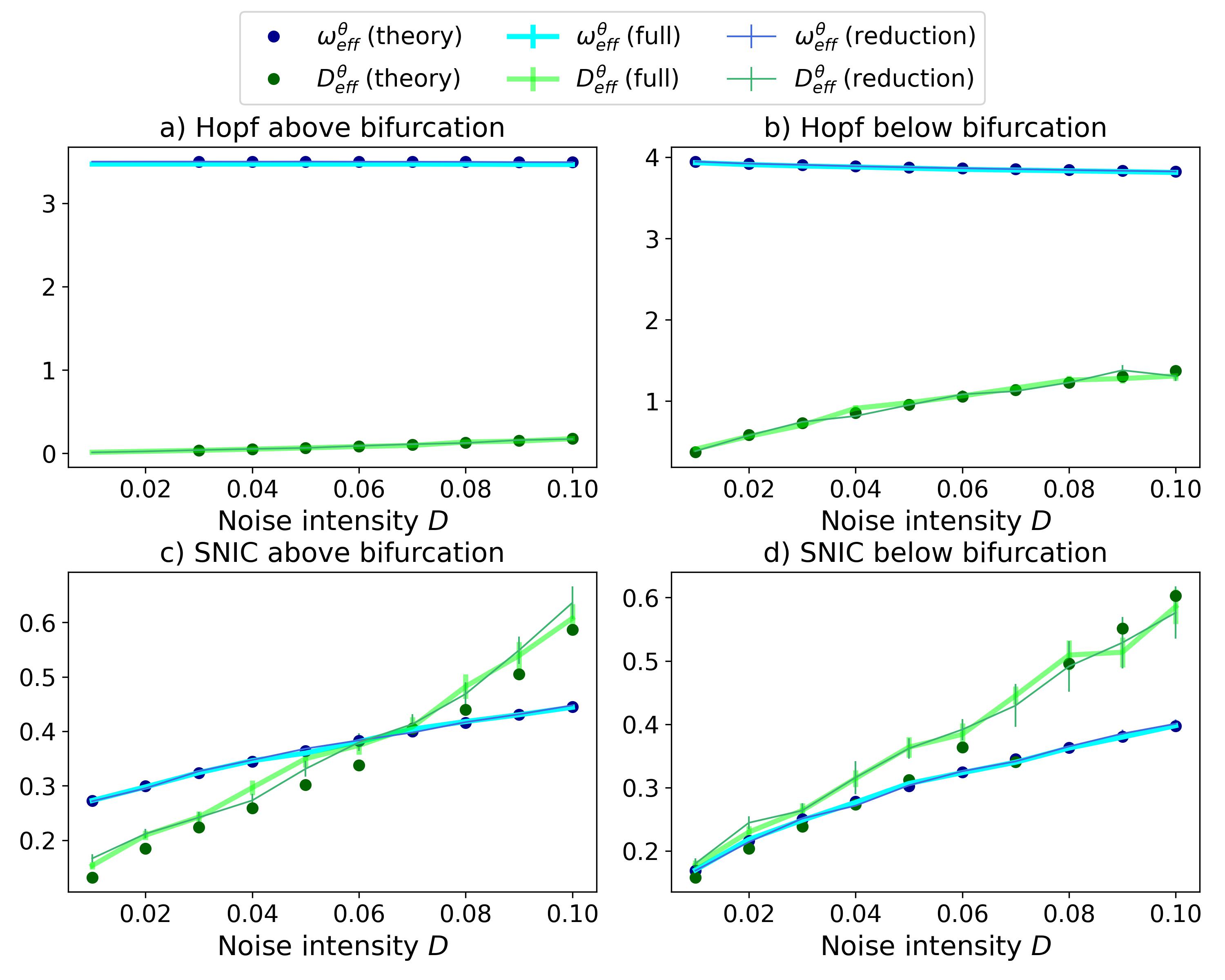}
\caption{\textbf{Long-term statistics of the MRT phase $\Theta(\mbx)$ as a function of the noise strength $D$.} (a) Hopf bifurcation in the LC case; (b) Hopf bifurcation in the focus case; (c) SNIC in the LC case; (d) SNIC in the excitable case. We compute each statistic for: the full phase equation \e{mrt-Dyn} (thick line), its phase reduction equation \e{ideal-phase-mrt} (narrow line) and the theoretical formulas \e{benjamin-formula-rot} and \e{benjamin-formula-dif} (dots). Error bars span one standard error.}
\label{fig:mrt-stat_test}
\end{figure}

\begin{figure}[t]
	\centering
	\includegraphics[width=0.48\textwidth]{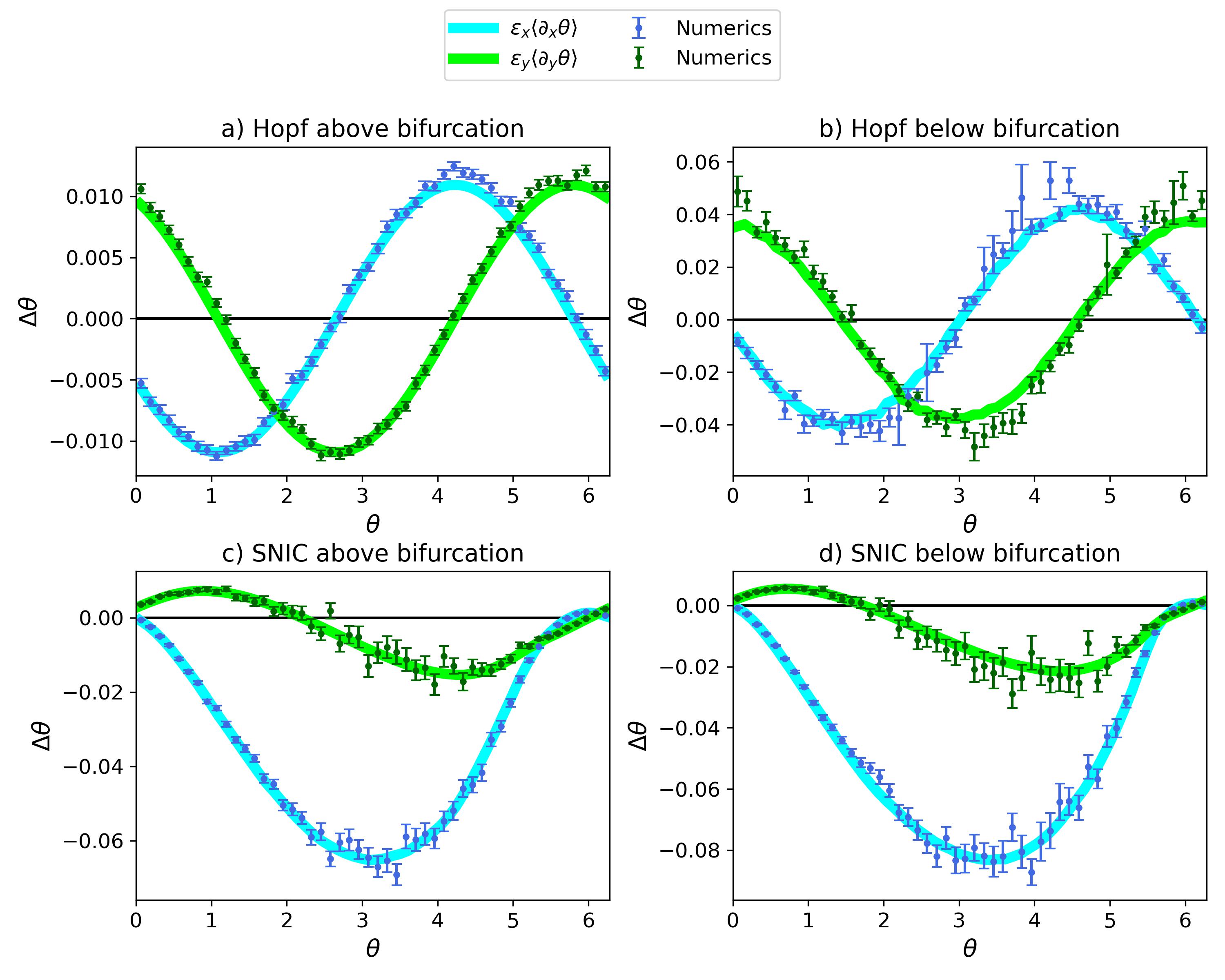}
\caption{\textbf{Averaged iPRCs for the MRT phase $\Theta(\mbx)$}. Blue - response to a pulse in the X direction (amplitude $\epsilon_x$);  Green - response to a pulse in the Y direction ($\epsilon_y$) a) Hopf above bifurcation b) Hopf below bifurcation c) SNIC above bifurcation d) SNIC below bifurcation. External level of noise used for all systems: $D= 0.01$; pulse amplitudes: $\epsilon_x = \epsilon_y = 0.01$. Error bars span one standard error.}
	\label{fig:MRT-aiPRC}
\end{figure}

\textit{MRT long-term statistics:} 
Following sec.\ref{sec:sec-5}, we compute the mean rotation rate $\omega^\theta_\text{eff}$ in \eqref{eq:long-term-drift} and the diffusion coefficient $D^\theta_\text{eff}$ in \eqref{eq:long-term-diffusion} by means of: (i) the full mapping $\Theta(\mbX(t))$ solution of \e{mrt-Dyn}; and (ii) the self contained reduction $\theta(t)$, solution of \e{ideal-phase-mrt}. 
Additionally, as was done in sec.\ref{sec:sec-5}, since the general phase reduction is a 1D SDE with periodic drift and noise coefficients, we use the results in \cite{LinSch02} to compute the mean rotation rate and the diffusion  coefficient by means of the theoretical expressions eqs.~\eqref{eq:benjamin-formula-rot} and \eqref{eq:benjamin-formula-dif}. Results are shown in Fig.~\ref{fig:mrt-stat_test}. As for the stochastic asymptotic phase results in Fig.~\ref{fig:stat_test}, we find a very good level of agreement between the values of $\omega^\theta_\text{eff}$ and $D^\theta_\text{eff}$ for the full and the reduced phase dynamics.
\textit{MRT averaged iPRC:} Following sec.\ref{sec:sec-6}, we compute the \textit{averaged infinitesimal Phase Response Curve} (aiPRC) for the MRT phase $\nabla \Theta(\mbx)$ as an average across realizations \eqref{eq:aiPRC_trajectories}. Results are shown in Fig.~\ref{fig:MRT-aiPRC}, and present a good agreement between predicted and measured responses.

\section{Morris-Lecar Parameters}\label{app:ML-param}
The parameters used for the 3D Morris-Lecar model are given in Table~\ref{table-ml}:
\begin{table}[h]
\begin{ruledtabular}
	\begin{tabular}{cc}
        $m_\infty(V)$: $0.5\Big[1 + \tanh{(\frac{V-\beta_m}{\gamma_m})} \Big]$ & $\beta_m$: -1.2 mV, $\gamma_m$: 18 mV\\
        $Y_\infty(V)$: $0.5\Big[1 + \tanh{(\frac{V-\beta_Y}{\gamma_Y})} \Big]$ & $\tau_Y(V)$: $1/\cosh(\frac{V - \beta_Y}{2\gamma_Y})$\\
        $\beta_Y$: -10 mV, $\gamma_Y$: 10 mV & $\phi_Y$: 0.15\\
        $Z_\infty(V)$: $0.5\Big[1 + \tanh{(\frac{V-\beta_Z}{\gamma_Z})} \Big]$ & $\tau_Z(V)$: $1/\cosh(\frac{V - \beta_Z}{2\gamma_Z})$\\
        $\beta_Z$: -21 mV, $\gamma_Z$: 15 mV & $\phi_Z$: 0.5 \\
        $E_\text{Na}$: 50 mV & $E_\text{K}$: -100 mV \\$E_\text{L}$: -70 mV & $E_\text{sub}$: 50 mV \\
        $g_\text{fast}$: 20 mS/cm$^2$ & $g_\text{K, dr}$: 20 mS/cm$^2$ \\
        $g_\text{sub}$: 2 mS/cm$^2$ & $g_\text{L}$: 2 mS/cm$^2$ \\
        C: 1 $\mu$F/cm$^2$ & $I_\text{ext}$: 29 $\mu$A/cm$^2$
        
	\end{tabular}
 \caption{Parameters of the 3D Morris-Lecar model in \eqref{eq:3DML}.}
\label{table-ml}
\end{ruledtabular}
\end{table}

\newpage

\begin{comment}
\section{gEDMD}

Following \cite{KluNus20}, we provide details on the gEDMD procedure that was used here. 

For stochastic differential equations
\begin{equation*}
\diff \mbX = \mbf(\mbX) \diff t + \mbg(\mbx)\diff \mbW_t,
\end{equation*}
$\mbX \in \mathbb{R}^d$, the infinitesimal generator of the stochastic Koopman operator acting on any twice continuously differentiable function $F$ is defined as
\begin{align*}
    \LLd[F(\mbx)] &= \diff F(\mbx) = \mbf(\mbx) \cdot \nabla F(\mbx) + \frac{1}{2} \mbg(\mbx)\mbg (\mbx)^\top: \nabla^2 F(\mbx)\\
    &= \sum_{i=1}^d f_i (\mbx)\frac{\partial F(\mbx)}{\partial x_i} + \frac{1}{2}\sum_{i=1}^d \sum_{j=1}^d (\mbg(\mbx)\mbg(\mbx)^\top)_{ij} \frac{\partial F(\mbx)}{\partial x_i \partial x_j}.
\end{align*}
gEDMD uses a set of training data $\{\mbX_l\}^m_{l=1}$, the drift and diffusion functions $\mbf$ and $\mbg$, and a library of functions $\{F_j\}^k_{j=1}$ to build $L$, the least-square approximation of $\LLd$:
\[L = \Big( \diff \mbF_X \mbF_X^\dagger \Big)^\top = \Big((\diff \mbF_X \mbF_X^\top)(\mbF_X \mbF_X^\top)^\dagger \Big)^\top.\]
where
\[\mbF_X = \begin{bmatrix}
    F_1(\mbx_1) & \cdots & F_1(\mbx_m) \\
    \vdots & \ddots & \vdots \\
    F_k(\mbx_1) & \cdots & F_k(\mbx_m)
\end{bmatrix},\]
\[\diff \mbF_X = \begin{bmatrix}
    \diff F_1(\mbx_1) & \cdots & \diff F_1(\mbx_m) \\
    \vdots & \ddots & \vdots \\
    \diff F_k(\mbx_1) & \cdots & \diff F_k(\mbx_m)
\end{bmatrix}.\]
In the worked example, we used a polynomial basis as the library.
    
\end{comment}

\bibliography{biblio}% Produces the bibliography via BibTeX.

\end{document}